\documentclass[
 reprint,
 amsmath,amssymb,
 aps,
]{revtex4-2}

\usepackage{graphicx}
\usepackage{dcolumn}
\usepackage{bm}
\usepackage{dsfont}
\usepackage{amsthm,amsmath,amssymb}
\usepackage{exscale}
\usepackage{relsize}
\usepackage{xcolor}
\usepackage{appendix}

\usepackage{mathrsfs}
\begin{document}

\preprint{APS/123-QED}

\title{Composite picosecond control of atomic state through a nanofiber interface}

\author{Yudi Ma}
\email{ydma18@fudan.edu.cn}
\author{Ruijuan Liu}
\author{Lingjing Ji}
\author{Liyang Qiu}
\author{Saijun Wu}
\email{saijunwu@fudan.edu.cn}

\affiliation{Department of Physics, State Key Laboratory of Surface Physics and Key
Laboratory of Micro and Nano Photonic Structures (Ministry of Education),
Fudan University, Shanghai 200433, China.}

\author{Dianqiang Su}
\author{Yanting Zhao}
\email{zhaoyt@sxu.edu.cn}
\affiliation{$^1$State Key Laboratory of Quantum Optics and Quantum Optics Devices, Institute of Laser Spectroscopy, Shanxi University, Taiyuan 030006, China.\\
$^2$Collaborative Innovation Center of Extreme Optics, Shanxi University, Taiyuan 030006, China
}

\author{Ni Yao}
\author{Wei Fang}
\email{wfang08@zju.edu.cn}

\affiliation{Interdisciplinary Center for Quantum Information, State Key Laboratory of Modern Optical Instrumentation,
College of Optical Science and Engineering, Zhejiang University, Hangzhou 310027, China
}

\date{today}

\begin{abstract}
Atoms are ideal quantum sensors and quantum light emitters. Interfacing atoms with nanophotonic devices promises novel nanoscale sensing and quantum optical functionalities.  But precise optical control of atomic states in these devices is challenged by the spatially varying light-atom coupling strength, generic to nanophotonic. We demonstrate numerically that despite the inhomogenuity, composite picosecond optical pulses with optimally tailored phases are able to evanescently control the atomic electric dipole transitions nearly perfectly, with $f>99\%$ fidelity across large enough volumes for {\it e.g.} controlling cold atoms confined in near-field optical lattices. Our proposal is followed by a proof-of-principle demonstration with a $^{85}$Rb vapor -- optical nanofiber interface, where the excitation by an $N=3$ sequence of guided picosecond D1 control reduces the absorption of a co-guided nanosecond D2 probe by up to $\sim70\%$. The close-to-ideal performance is corroborated by comparing the absorption data across the parameter space with first-principle modeling of the mesoscopic atomic vapor response. Extension of the composite technique to $N\geq 5$ appears highly feasible to support arbitrary local control of atomic dipoles with exquisite precision. This unprecedented ability would allow error-resilient atomic spectroscopy and open up novel nonlinear quantum optical research with atom-nanophotonic interfaces. 
\end{abstract}

\maketitle

\section{Introduction}
Atoms are ideal quantum sensors and quantum light emitters. With atomic spectroscopy, the centers and linewidths of optical transitions are precisely measured for probing external potentials and unknown interactions~\cite{Sansonetti2011,Lu2013,Fuchs2018,Patterson2018,Solano2019c,Peyrot2019a,Hummer2021}. In quantum optics, collective absorption and emission by ensemble of atoms are controlled for applications ranging from nonlinear frequency generation~\cite{Eikema1999,Gabrielse2018} to quantum information processing~\cite{Duan2001,Srakaew2023}. Interfacing atoms with nanophotonic devices promises exciting perspectives for atomic spectroscopy at nanoscales~\cite{Fuchs2018,Patterson2018,Solano2019c,Peyrot2019a,Hummer2021} and  for implementing efficient nonlinear optics~\cite{Spillane2008,Hendrickson2010,Venkataraman2011,Finkelstein2021}. Furthermore, with the development of laser cooling technology~\cite{MetcalfBook}, atoms are cooled and optically trapped in the near field of nanophotonic structures~\cite{Vetsch2010,Meng2018,Su2019, Samutpraphoot2020} to support many-body interaction mediated by exchange of confined photons~\cite{Chang2018,Solano2019b, Corzo2019, Samutpraphoot2020, Pennetta2022,Pennetta2022b}. Quite obviously, in all these scenarios that exploit  the strong, electric dipole light-atom interaction, it would be highly useful if one can precisely control the 2-level ``optical spins'' defined on the strong transitions, similar to those achieved in nuclear magnetic resonance (NMR)~\cite{Garwood2001a,Levitt1986} or with 2-photon narrow transitions~\cite{Levine2018,Fluhmann2019}. For example, such ability would support efficient excitation in nanoscale atomic spectroscopy~\cite{Fuchs2018,Patterson2018,Solano2019c,Peyrot2019a,Hummer2021} while  suppressing the light shift and broadening~\cite{Sanner2018,Yudin2020}. The precise optical spin control would enable arbitrary, high contrast optical modulation in nonlinear spectroscopy~\cite{Spillane2008,Hendrickson2010,Venkataraman2011,Finkelstein2021} to improve the sensitivity to time-dependent perturbations~\cite{Yuge2011,Wang2022e}. For atomic ensembles, precise control of the atomic dipoles can be useful for steering collective couplings between atoms and the guided light~\cite{Solano2019b, Corzo2019, Pennetta2022,Pennetta2022b}, or even to coherently suppress the couplings~\cite{Scully2015,He2020a} for accessing the subradiant physics in the confined geometry~\cite{Noh2017, Albrecht2017a,  Kornovan2019, Zhang2019, Zhang2020b, Prasad2020,  Poshakinskiy2021b}. However, unlike NMR or narrow line controls, high precision optical control of electric dipole strong transitions is itself an outstanding challenge~\cite{Ma2020}. The technical difficulty is amplified at nanophotonic interfaces where a uniform optical control seems prohibited by a none-uniform light intensity and polarization distribution.


\begin{figure*}
    \centering
    \includegraphics[width=1\textwidth]{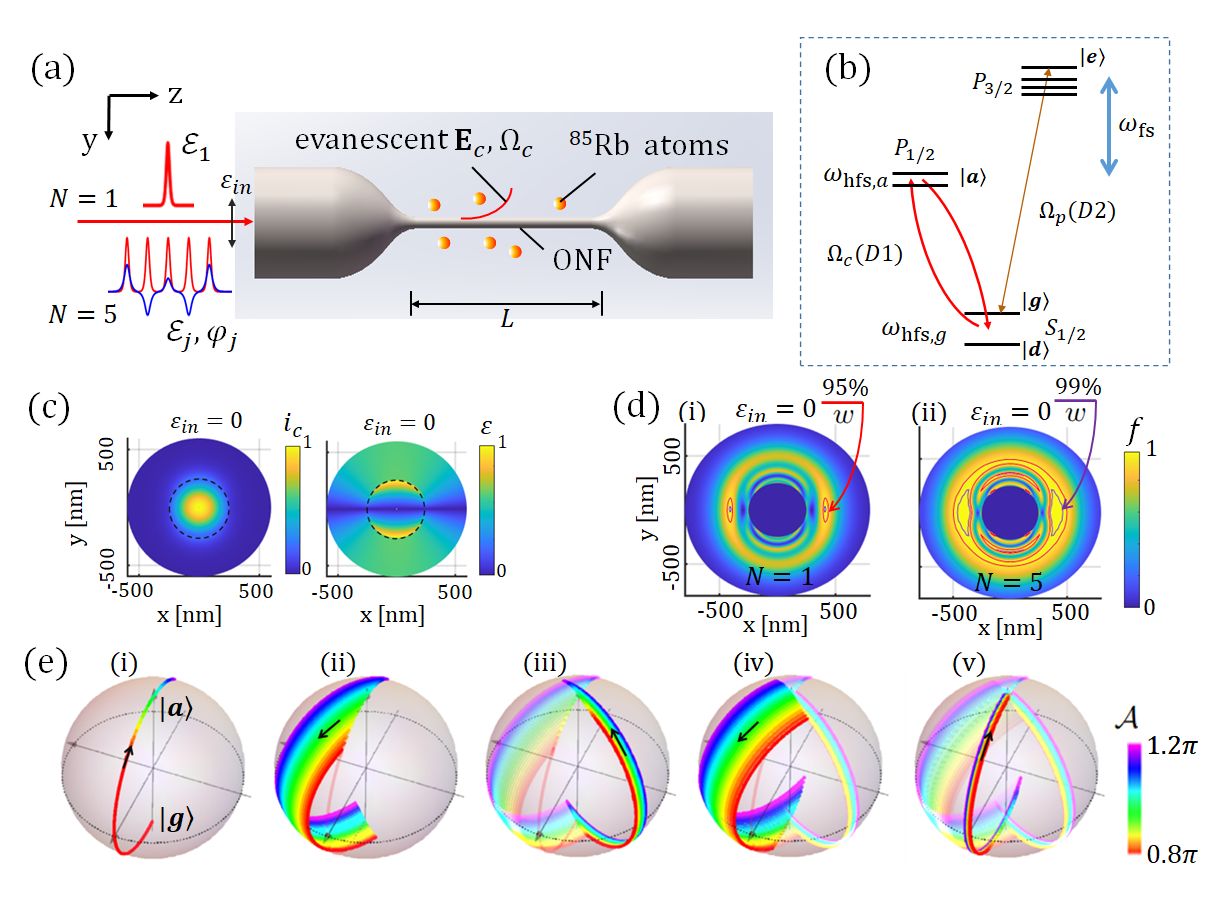}
    \caption{ (a): Schematic setup for composite picosecond control of atomic states through an atom-optical nanofiber (ONF) interface, with exemplary $N=1$ and $N=5$ picosecond ${\bf E}_c$ sequence sent through ONF to resonantly drive the D1 transition of free $^{85}$Rb in the near field. The atomic population in the $F=3$ ground-state hyperfine level ($|g\rangle$) can be monitored by a co-guided D2 probe (See Fig.~\ref{fig:probe}a).  The red (blue) lines stand for the intensity $|{\bf E_c}|^2$ (real envelope ${\rm Re}(E_c)$) of the composite pulse. (b): The $^{85}$Rb level diagram and optical  coupling scheme in this work. (c): $i_c(x,y)$ and $\varepsilon(x,y)$ distributions near the $d=500$~nm ONF for the HE$_{11}^y$-coupled control pulses ($\varepsilon_{\rm in}=0$, also see Appendix~\ref{sec:circular}). The normalized intensity $i_c(x,y)=|{\bf E}_c(x,y)|^2/|{\bf E}_c({\bf 0})|^2$ decays radially with a characteristic field decay length $\xi\approx 1/\sqrt{\beta_{\rm c}^2-k_c^2}$ of 200~nm ($\beta_{\rm c}\approx 1.2 k_c$ is the propagation constant. $k_c=2\pi/795$~nm). The local ellipticity is defined as $\varepsilon(x,y)=|{\bf E}_c^*\times {\bf E}_c|/|{\bf E}_c|^2$. (d): Population inversion by the highly inhomogenuous HE$_{11}^y$ coupled $N=1$ and $N=5$ control pulses. For  stationary atoms~\cite{Vetsch2010,Meng2018,Su2019},  $f=\rho_{aa}(\tau_c)>99\%$ efficiency can be achieved locally by optimizing the relative phases $\{\varphi_j\}$ of the $N=5$ composite pulse. See the $f=95\%$ (red curves), $f=99\%$ (purple curves, with $w\approx$110~nm width) contours in the Fig.~(d,ii) simulation, with $\{\varphi_{j}\}=\{0,5\pi/6,2\pi/6,5\pi/6,0\}$ according to ref.~\cite{Genov2014}.  For comparison, for the single pulse ``$\pi$'' excitation, even the $f=95\%$ contour in Fig.~(d,i) is limited to a $w\approx$40~nm width. The $f=99\%$ contour is hardly visible, which completely vanishes with $\varepsilon_{\rm in}=0.1$ (not shown). (e): Bloch sphere representation of the simplified 2-level $|g\rangle-|a\rangle$ dynamics subjected to $N=1$ single pulse (represented by Fig.~(e,i) alone), and $N=5$ composite pulses~\cite{Genov2014} (Trajectories driven by each of the $N=5$ pulses are highlighted by Fig.~(e,i-v) respectively.), with equal pulse area $\mathcal{A}_j=\mathcal{A}\in(0.8\pi,1.2\pi)$. }
\label{fig:Setup}
\end{figure*}


We propose to achieve precise nanophotonic control of atomic states by implementing a class of composite control schemes~\cite{Genov2014, Low2016} with picosecond optical excitations~\cite{Ma2020}. We demonstrate numerically that despite the near-field inhomogenuity, electric dipole transitions of alkaline atoms can be controlled nearly perfectly, with $f>99\%$ fidelity across large enough volumes for near-field samples such as cold atoms confined in near-field optical lattices~\cite{Vetsch2010,Meng2018,Su2019}. 
Our proposal is followed by a proof-of-principle demonstration of the composite picosecond control where an $N=3$ sequence of guided picosecond pulses robustly invert the D1 population of free-flying $^{85}$Rb atoms trespassing an ONF interface. The effect of population inversion is monitored by the transiently enhanced transmission of a co-guided nanosecond probe pulse. By optimizing the relative phases among the sequence of transform-limited picosecond pulses~\cite{Ma2020}, a $\sim70\%$ enhancement of the probe transmission is observed (even though microscopically this enhancement is after all the near-field thermal averaging). The measurements are compared with first-principle modeling of light-atom interaction. The agreements across the parameter space of the composite control strongly suggest that our $N=3$ sequence reaches a close-to-ideal performance~\cite{Genov2014}. 

Our work paves a practical pathway toward large $N$, composite optical control of strong transitions for atoms confined in the near field~\cite{Vetsch2010,Meng2018,Su2019}, with an exquisite precision similar to those achieved in magnetic resonances~\cite{Genov2014, Rong2015}. The nearly perfect dipole control could open up a variety of research opportunities with nanophotonic interfaces, such as for error-resilient near-field spectroscopic sensing~\cite{Fuchs2018,Patterson2018,Solano2019c,Peyrot2019a,Hummer2021} with composite pulses~\cite{Sanner2018,Yudin2020, Yuge2011,Wang2022e}, for studying nanophotonic nonlinear optics~\cite{Spillane2008,Hendrickson2010,Venkataraman2011,Finkelstein2021} with unprecedented control precision and flexibility,
and for developing novel quantum optical devices by harnessing the collective couplings between the atomic ensemble and the nanophotonically confined light~\cite{Scully2015,He2020a,Chang2018}.

%


%

In the following the paper is structured into three sections. First, in Sec.~\ref{sec:scheme}, we outline our theoretical proposal for the picosecond composite control at nanophotonic interfaces. Using the D lines of alkaline atoms and a composite population inversion scheme~\cite{Genov2014} as example, we numerically demonstrate that composite techniques can be implemented to evanescently control atomic states in a nearly perfect manner, over a large volume in the near field. We highlight the control power efficiency enabled by the nanophotonic optical confinement, to potentially support $N\sim 100$ sub-pulses for realizing highly sophisticated maneuvers~\cite{Low2016} using moderately strong pulses from a mode-locked laser~\cite{Ma2020}. Next, in Sec.~\ref{sec:exp}, we detail our experimental measurements with a thermal vapor-ONF interface to confirm the practicality of this proposal. We present first-principle numerical simulations to compare with the measurements, from which we infer the performance for an $N=3$ robust population inversion at the interface. The technical limitation to the $N\leq 3$ control in this demonstration is discussed in Sec.~\ref{sec:discussion}A. We conclude this paper in Sec.~\ref{sec:discussion}B with an outlook into future research opportunities opened up by the picosecond composite control technique.


\section{Composite picosecond control at nanophotonic interface}\label{sec:scheme}

Precise control of strong optical transitions can be achieved at a carefully chosen time scale. As in Fig.~\ref{fig:Setup}b, we consider arbitrary control of a $|g\rangle-|a\rangle$ ``optical spin'' defined on the D1 line of an alkaline atom, using a control pulse with Rabi frequency $\Omega_c$ and duration $\tau_c$. Clearly, $\tau_c\gg 1/\omega_{\rm fs}$ to resolve the fine-structure-split optical lines, and $\tau_c\ll 1/\Gamma_a$  to avoid spontaneous emission ($\Gamma_a$ the natural linewidth of the excited $|a\rangle$ states), are required for achieving the coherent 2-level ``spin'' control. Meanwhile, $\tau_c\ll 1/\omega_{{\rm hfs},g},  1/\omega_{{\rm hfs},a}$ is preferred so that the multi-level dynamics associated with hyperfine Raman excitations~\cite{Campbell2010,Qiu2022} are relatively easy to manage (Appendix~\ref{sec:composite}). The effect of the D1 control can be monitored by a weak D2 probe, if needed. The laser of choice for the D1 control is a picosecond laser~\cite{Long2019, Hussain2020,Ma2020}. The $\sim100$~GHz bandwidth is narrow enough to resolve atomic lines with multi-THz separations, wide enough to cover the typical hyperfine splitting at the GHz level, and support quick enough pulsed control to avoid radiation damping. Furthermore, a moderate Rabi frequency $\Omega_{\rm c}$ at the 100~GHz level is sufficient for the picosecond control. The moderate strength helps avoiding {\it e.g.}, photo-ionization during multiple controls~\cite{Zhdanovich2008}, and practically enable efficient implementation with low-energy pulses. 

The light confinement enables efficient optical control at nanophotonic interfaces~\cite{Piatkowski2016, Dombi2020, Spillane2008,Hendrickson2010,Venkataraman2011,Finkelstein2021}. We  consider the specific example of the ONF interface illustrated in Fig.~\ref{fig:Setup}, where the D1 transition of the proximate $^{85}$Rb atoms is controlled by transform-limited picosecond pulses of $\lambda_{\rm c}=795~$nm light guided through a step-index silica nanofiber~\cite{Su2019} in the fundamental HE$_{11}$ mode (Appendix~\ref{sec:circular})~\cite{Tong2004}. At a $d=500$~nm fiber diameter, about $20\%$ of the light power propagates evanescently in vacuum to interact with the atoms. The electric dipole control is remarkably efficient. 
A resonant picosecond pulse energy of merely $\mathcal{E}_{1}(\pi)\sim 1~{\rm pJ\cdot ps}/\tau_c$ is sufficient for the $|g\rangle-|a\rangle$ Rabi frequency $\Omega_c\propto E_c$ (Eq.~(\ref{eq:rRbi})) to reach $\pi/\tau_c$ near the ONF surface, for locally driving a ``$\pi$'' inversion within $\tau_c$.  Here $E_c({\bf r},t)$ is the amplitude of the complex electric field envelop function ${\bf E}_c({\bf r},t)$ in the near field.


We are interested in the actual quality of the D1 inversion. Unfortunately, the strong optical confinement (Fig.~\ref{fig:Setup}c) is associated with light intensity and polarization inhomogenuities~\cite{Tong2004,Lodahl2017,Jones2020} to prevent a simple pulse from achieving a uniform inversion in the near field, particularly in presence of the $|g\rangle-|d\rangle$ Raman couplings~\cite{Campbell2010}. Nevertheless, it is important to note that when the control pulse is short enough, $\tau_c\ll 1/\omega_{{\rm hfs},g}, 1/\omega_{{\rm hfs},a}$, the hyperfine $|g\rangle-|a\rangle$ transition in the D1 example reduces to a $J_g=1/2\leftrightarrow J_a=1/2$ transition (Appendix~\ref{sec:JI}). By choosing the quantization axis along the local helicity axis ${\bf e}_h={\bf E}_c^*\times {\bf E}_c/|{\bf E}_c|^2$, the D1 excitation is decomposed into 2-level $\sigma^{\pm}$ couplings, with a relative coupling strength determined by the local ellipticity of light $\varepsilon(x,y)=|{\bf e}_h|$ (Eq.~(\ref{eq:omJI})). For the limiting case of a linearly polarized field (with quantization axis perpendicular to ${\bf E}_c$), the coupling strengths become degenerate. 
One thus expects nearly perfect inversion efficiency insensitive to initial $|g\rangle$ Zeeman sub-levels (Fig.~\ref{fig:coupling}), at specific intensities $|{\bf E}_c|^2$. This is confirmed with our full-level simulations (Appendix~\ref{sec:D1})~\cite{Sievers2015,Bruce2017,Qiu2022}, illustrated in Fig.~\ref{fig:Setup}(d,i) with an example 2D distribution of $|g\rangle - |a\rangle$ inversion efficiency $f(x,y)$ for stationary, initially unpolarized atoms in $|g\rangle$ (see Eq.~(\ref{eq:f})). Here, for a control pulse ${\bf E}_c$ in the HE$_{11}^y$ mode with an incident ellipticity $\varepsilon_{\rm in}=0$ (Fig.~\ref{fig:Setup}a), the near-field ellipticity is minimized near $y=0$ (Fig.~\ref{fig:Setup}c), $\varepsilon(x,y)\ll 1$ to support the nearly perfect inversion within $\tau_c=6$~ps time, with $\mathcal{E}_1=0.4~$pJ here. However, the inversion is not robust. A nanoscale shift of atomic position can degrade the performance substantially. Furthermore, the $f=99\%$ contour (which is barely visible in Fig.~\ref{fig:Setup}(d,i)) vanishes at a moderate $\varepsilon_{\rm in}=0.1$, as suggested by additional numerical simulations. We note such a slight change of the HE$_{11}$ polarization state~\cite{Tong2004} could easily be induced by the birefringence of a stressed fiber. 



To improve the control robustness, composite pulse sequence with $N$ sub-pulses can be programmed to achieve highly accurate control with uniform efficiency by exploiting the geometric phase of 2-level transitions~\cite{ICHIKAWA2012}. Such composite techniques are well developed in the research field of nuclear magnetic resonance~\cite{Levitt1986,Genov2014, Low2016}.
In particular, a sequence of $N$-pulses with equal amplitude $\{a_j\}$ and optimal phase $\{\varphi_j\}$, $j=1,...,N$ can be applied to invert the population of a 2-level spin in a manner which is highly resilient to the field strength and frequency detuning errors~\cite{Genov2014}. 

In the optical domain, the NMR-inspired composite technique can be exploited for driving ensemble of atoms illuminated by an inhomogeneous laser field, as well as for driving ensemble of transitions with different transition strengths (and slight different transition frequencies) for a same atom. The ONF-based atomic state control in this work exploits both aspects of the control technique (Appendix~\ref{sec:JI}). For the aforementioned reasons associated with quantum control timescales, the composite pulses need to be synthesized within picoseconds  with full waveform programmability, a requirement within a technical gap between the laser modulation technologies for continuous wave and ultrafast lasers~\cite{Gould2015}. In a recent work~\cite{Ma2020}, we developed a ``direct ${\bf k}$-space-to-time pulse shaping'' method~\cite{Emplit92, Leaird99, Mansuryan2011} to precisely generate $\tau_c\sim 100$~ps waveforms with tens to hundreds of GHz bandwidth, only limited by that of the seeding mode-locked laser. With the method, as detailed in Appendix~\ref{sec:generator}, an incident single pulse with energy $\mathcal{E}_0$ and a full pulse width at half maximum $\tau_0$ can be shaped into $j=1,...N$, $\tau_j=(j-1)\tau_{\rm d}$ even-delayed sub-pulses with energy $\{\mathcal{E}_j\}$ and phases $\{\varphi_j\}$, with fully programmable amplitudes and phases of the sub-pulses. Taking into account the sech$^2$-shaped soliton pulses from the mode-locked laser~\cite{sech2pulse}, we associate a total control duration $\tau_c=(N-1) \tau_{\rm d}+\tau$ to the composite pulse with $\tau=2\tau_0$ to characterize the control in the time domain. 

A particular example of robust atomic state control is illustrated in Fig.~\ref{fig:Setup}(d,ii) according to numerical simulation of $^{85}$Rb atom driven by the near field composite D1 excitation (Sec.~\ref{sec:model}). The robust control is best illustrated on a Bloch sphere first, as in Fig.~\ref{fig:Setup}e, which simplifies the hyperfine $|g\rangle-|a\rangle$ transition (Appendix~\ref{sec:JI}) with 2-level dynamics. The pulse areas are defined as $\mathcal{A}_j=|\int \Omega_j {\rm d}t|$ near the ONF surface, with $\Omega_j\propto \sqrt{\mathcal{E}_j}e^{i\varphi_j}$ to be the Rabi frequency of the sub-pulse $j$ (Eqs.~(\ref{eq:rRbi})(\ref{equ:inOut})). By optimizing the relative phases $\{\varphi_j\}$ among the sub-pulses of an $N=5$ equal-$\mathcal{E}_j$ sequence, according to ref.~\cite{Genov2014}, the seemingly redundant rotations of the state vector along the $\vec{\Omega}_j\equiv({\rm Re}(\Omega_j),{\rm Im}(\Omega_j),0)$ axes  (Fig.~\ref{fig:Setup}(e,i-v)) are phased in a way to cancel the inhomogenuous $\delta\mathcal{A}\in(-0.2\pi, 0.2\pi)$ deviation from $\mathcal{A}_j=\pi$. The final state inversion fidelity in Fig.~\ref{fig:Setup}(e,v) reaches $f> 99.97\%$~\cite{Genov2014}. This is in contrast to  the simple single-pulse control represented by Fig.~\ref{fig:Setup}(e,i) alone, which only reaches $f\approx 90\%$ on average due to the same $\delta \mathcal{A}$ broadening, hardly avoidable in the near field.

We now apply the $N=5$ optimal composite excitation~\cite{Genov2014} to the $^{85}$Rb D1 interaction in the near field. For the atom starting from an arbitrary Zeeman sub-level of $|g\rangle$ (Fig.~\ref{fig:coupling}) at $t=0$, with $\mathcal{E}_j=0.4$~pJ here, $f>99\%$ population inversion around $y=0$ is achieved across a substantial area where the near field is approximately linear (the purple contour in Fig.~\ref{fig:Setup}(d,ii)). The width $w\approx 110$~nm along the narrower $x-$direction with $\delta \mathcal{A}\in (-0.3\pi,0.3\pi)$ is substantially larger than the 1-pulse case, where even for the $95\%$ contour the width is only $w\approx 40$~nm (Fig.~\ref{fig:Setup}(d,i)). The total control time is $\tau_c=66$~ps in the Fig.~\ref{fig:Setup}(d,ii) simulation, with a $\tau_0=3~$ps pulse width. We note longer $\tau_c$ leads to reduced width for the $f=99\%$ contour along $y$, with the $w$ along $x$  unaffected, as long as $\tau_c\ll 1/\omega_{{\rm hfs},a}$ so that 2-photon Raman transitions by linear polarized excitations are suppressed~\cite{Happer1972} (Appendix~\ref{sec:JI}). The population inversion is robust against small incident polarization variations. In particular, the $f=99\%$ contour in Fig.~\ref{fig:Setup}(d,ii) is hardly changed when the incident polarization is increased to $\varepsilon_{\rm in}=0.1$ in the simulation. We thus expect robust implementation of the $N=5$ composite sequence to {\it e.g.} arrays of trapped atoms through the ONF interface~\cite{Vetsch2010,Meng2018,Su2019} to achieve ultra-precise population inversion.

Beyond simple population inversion, composite pulses can be shaped to implement more sophisticated controls, such as to geometrically shift the optical phases of the D2 dipoles with double D1 inversions~\cite{Scully2015, He2020a}, or even for driving arbitrary qubit gates~\cite{Low2016} on the optical transitions. For composite pulses with nearly equal amplitudes, the sub-pulse energy $\mathcal{E}_j < \mathcal{E}_0/N^2$ is limited by conservation of optical spectrum density. Nevertheless, with the $\mathcal{E}_0=N^2\mathcal{E}_1(\pi)$ scaling and  at a moderate pulse shaping efficiency~\cite{Ma2020}, the required input pulse energy $\mathcal{E}_0$ for sophisticated controls~\cite{Low2016,Sanner2018,Yudin2020, Yuge2011,Wang2022e} with up to $N=10^2$ sub-pulses is below 100~nJ. The required seeding pulse energy $\mathcal{E}_0$ is reduced further for nano-structures with better atom-light cooperativity~\cite{Ritter2018,Beguin2020,Leong2020,Finkelstein2021}. 



\section{A Proof-of-principle demonstration}\label{sec:exp}

The picosecond composite control technique is applicable not only to microscopically confined cold atoms, but also to thermal atoms that transiently trespass the nanophotonic interface. In this section, we exploit the thermal vapor-ONF interface to demonstrate the picosecond atomic state control. 

\begin{figure}[h!]
        \centering
       \includegraphics[width=0.48\textwidth]{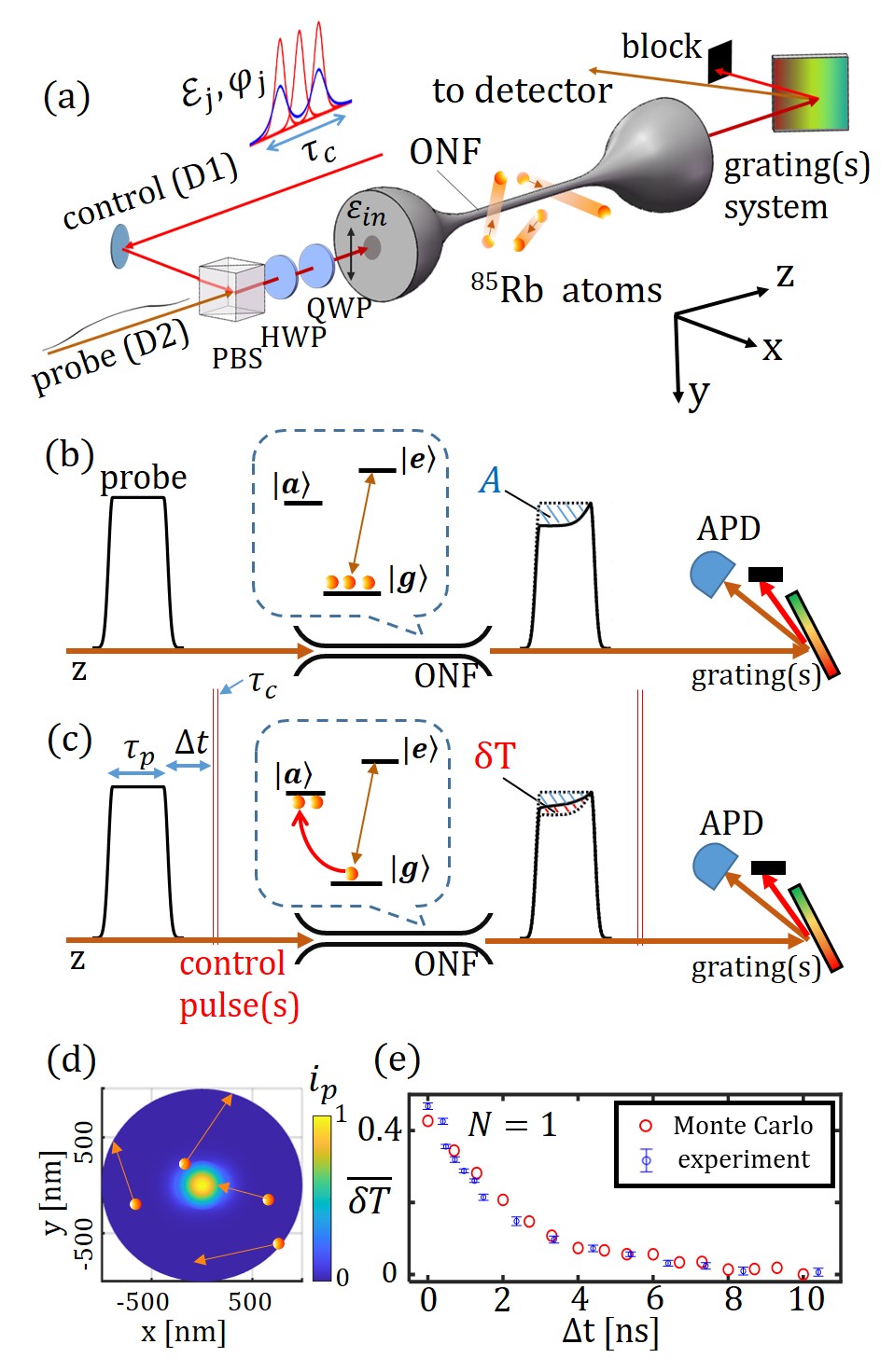}
        \caption{(a): Schematic of the control-probe scheme in this work to investigate the picosecond atomic state controlability at the ONF interface. The red and blue lines describe the composite picosecond pulse envelop, similar to Fig.~\ref{fig:Setup}a but with $N=3$ here. HWP: half-wave plate. QWP: quarter-wave plate. (b): A $\tau_p=2$~ns probe pulse resonant to the D2 $|g\rangle-|e\rangle$ transition is evanescently attenuated through the thermal vapour-ONF interface, leading to reduced transmission by  $A$ (an integration of the blue shaded area).  See Fig.~\ref{fig:Setup}b for the full level diagram including $|d\rangle$. (c): By firing the D1 control pulses ($|g\rangle-|a\rangle$) to deplete the ground state population $\rho_{gg}$, the $\Delta t$-delayed D2 probe attenuation is reduced by $\delta T$ (an integration of the red shaded area). The sub-$\tau_p$ $A$ and $\delta T$ transients are explained in Appendix~\ref{sec:mc}. The $\overline{\delta T}\equiv \delta T/A$ ratio reflects reduced $|g\rangle$ population in the near field.  (d):  Schematic illustration of thermal atoms trespassing the ONF evanescent field. Here $i_{\rm p}(x,y)=|{\bf E}_{\rm p}(x,y)|^2/|{\bf E}_{\rm p}({\bf 0})|^2$.
        (e): Experimental and simulated (Eq.~(\ref{eq:deltaTt})) transient transmission $\overline{\delta T}$ vs $\Delta t$, for $N=1$ single D1 pulse control (See Appendix~\ref{sec:calib}.). The error bars represent the statistical uncertainties from 10 repeated measurements.} 
    \label{fig:probe}
\end{figure}

\subsection{Measurement principle}\label{sec:expPricipal}
The experimental setup illustrated schematically in
Fig.~\ref{fig:probe}a follows a control-probe strategy. A nanosecond probe pulse with duration $\tau_{\rm p}$, resonant to the $|g\rangle-|e\rangle$ D2 transition of $^{85}$Rb, is combined with the picosecond $|g\rangle-|a\rangle$ D1 control pulses with duration  $\tau_{\rm c}$ through a polarization-dependent beamsplitter and sent to the ONF interface. The evanescent coupling between the ONF-guided probe pulse with the thermal atoms surrounding ONF leads to the probe attenuation.  As in Fig.~\ref{fig:probe}b, we denote the fractional attenuation to the probe transmission $T$ by the atomic absorption as $A=\Delta T/T$. In presence of the resonant D1 control pulses that also interact with the atoms evanescently (Fig.~\ref{fig:probe}c), the ground state population $\rho_{gg}$ in the near field is depleted, leading to transiently enhanced transmission $\delta T=-\delta A$.


Optically, our thermal vapor-ONF system is well within a  mesoscopic regime featuring transient spectral broadening and non-local responses~\cite{Peyrot2019}. 
In particular, at a ${\rm T}=360$~K temperature to be discussed shortly, the $^{85}$Rb atoms has a thermal velocity of $v_{\rm T}=\sqrt{k_{\rm B} {\rm T}/M}\approx 190$~m/sec ($M$ is the atomic mass). For the ONF-guided probe with a propagation constant $\beta_{\rm p}\approx 1.2 k_{\rm p}$~\cite{Tong2004} ($k_{\rm p}=2\pi/780$~nm), the Doppler linewidth~\cite{Hughes2018}   $\sqrt{2{\rm ln}2} \beta_{\rm p}v_{\rm T}/\pi\sim 700$~MHz is broadened to $\sim 800$~MHz according to absorption spectroscopy, which, as unveiled by simulation (Appendix~\ref{sec:mc}), is due to near-field atomic motion associated with the nanosecond $\delta t=\xi_{\rm p}/v_{\rm T}$ ($\xi_{\rm p}=1/\sqrt{\beta_{\rm p}^2-k_{\rm p}^2}=200~$nm is the probe field decay length. The control field decay length $\xi$ is similarly defined, see Fig.~\ref{fig:Setup}.). As schematically illustrated in Fig.~\ref{fig:probe}(b,c),  for the pulsed control and probe, we expect both $A$ and $\delta T$ to vary rapidly within a nanosecond due to the Doppler broadening and the $x-y$ motion (Appendix~\ref{sec:mc}). Here, we record the $\tau_{\rm p}$-integrated $\delta T$ and $A$ to obtain the normalized transient transmission,
\begin{equation}
\overline{\delta T}\equiv \delta T/A.\label{eq:deltaT}
\end{equation}
Intuitively, $\overline{\delta T}$ is the fractional reduction of the light scattering power during $\tau_{\rm p}$ by the atoms surrounding the ONF, due to the excitation by the control pulse that transfers the atomic population away from the state $|g\rangle$. As detailed in Sec.~\ref{sec:model}, $\overline{\delta T}$ is compared with a numerical model to infer the ground-state depletion efficiency (also see Eqs.~(\ref{eq:ga})(\ref{eq:rhougg}))
\begin{equation}
    f_g({\bf r})\equiv \Delta \rho_{gg}({\bf r},\tau_c)/\rho_{gg}^{(0)},\label{eq:fg}
\end{equation}
as well as the $|g\rangle-|a\rangle$ population inversion efficiency 
\begin{equation}
    f({\bf r})\equiv \rho_{aa}({\bf r}, \tau_c)/\rho_{gg}^{(0)},\label{eq:f}
\end{equation} 
by the picosecond control applied during $0\leq t\leq \tau_c$.  Here the $\rho_{gg}^{(0)}$ at $t=0$ and $\rho_{aa}{(\tau_c)}$ at $t=\tau_c$ are the atomic population summed over the Zeeman and hyperfine sublevels. We note that in presence of the hyperfine $|d\rangle$-level (Fig.~\ref{fig:Setup}b) and 2-photon picosecond Raman couplings~\cite{Campbell2010}, we generally expect $f({\bf r})\neq f_g({\bf r})$.







\subsection{Experimental setup}
Experimentally, the rubidium partial pressure at the nanofiber location is maintained around $p=10^{-5}$~Torr by electronically heating a dispenser $\sim$30~cm away in the vacuum. Depending on the actual vapor pressure during different periods of this project, attenuation of $A\approx 0.15\sim 0.2$ to the ONF transmission is obtained for the probe locked to the F=3-F$'$=4 D2 hyperfine transition of $^{85}$Rb. A heater attached to the ONF mount helps to maintain a 90$^{\circ}$C or ${\rm T}=360$~K temperature to suppress condensation of Rb atoms to the ONF surface. The probe beam is pulsed with a duration $\tau_{\rm p}=2$~ns to match the transiently broadened optical response of atoms to be detailed shortly, and is sent to ONF
after a $\Delta t$ delay relative to the picosecond control (Fig.~\ref{fig:probe}c). With a pair of computer-motorized half and quarter waveplates (Appendix~\ref{sec:pol}), the adjustments of the polarization states of the orthogonal HE$_{11}$ modes for the control and probe pulses are computerized. To separate the probe pulse from the pump background in a polarization-independent manner at the ONF output, we split the combined beams with another PBS (not in Fig.~\ref{fig:probe}a) to filter each path with a holographic grating, both at the ``p'' polarization with $\sim 70\%$ diffraction efficiency. The grating-filtered probe signals are then overlapped to an avalanche photodiode (APD) module (Hamamatsu C5658, 1~GHz detection bandwidth). An additional interference filter to bandpass the 780~nm probe is inserted to fully remove the 795~nm control photon background. 

As detailed in Appendix~\ref{sec:electronics}, the nanosecond APD output is analyzed by a home-made analogue signal integrator which compares $10^7$ interleaved measurements with and without the control pulses in two seconds. The high speed differential measurements lead to quality signals capable of resolving $\delta T$ at the $0.1\%$ level, even though the probe power is kept at $P_{\rm p}\approx 10$~nW to avoid saturating the D2 absorption. Normalized transient transmission $\overline{\delta T}(\Delta t, \{\mathcal{E}_j,\varphi_j\})$ is recorded for various composite pulse sequence at certain $\Delta t$ delay. The ``steady-state'' absorption $A$ for the $\overline{\delta T}$ normalization is monitored instead with a slow photo-multiplier tube (Hamamatsu CR131), by comparing the resonant transmission with the off-resonant values during frequency scans of the D2-laser (See Appendix~\ref{sec:calib} discussions). Notice here and in the following $A$ and $\delta T$ are normalized by the generic ONF transmission $T$ estimated to be better than $90\%$~\cite{Su2019}.


We verify the transient broadening picture with a simple delayed-probe experiment. As in Fig.~\ref{fig:probe}e, while a $\tau_c=2\tau_0=24$~picosecond D1 control pulse (Appendix~\ref{sec:generator}) with strong enough $\mathcal{E}_1\approx 0.5$~pJ is able to reduce the atomic absorption by $\overline{\delta T}\approx 50\%$ at $\Delta t=0$ (Also see Sec.~\ref{sec:sat}, Appendix~\ref{sec:mc},\ref{sec:calib}), the transmission recovers rapidly with a time constant as short as $2$~ns. After the picosecond excitation, atoms at the vicinity of ONF transferred to either the excited $|a\rangle$ or the other $F=2$ hyperfine ground states $|d\rangle$ (Fig.~\ref{fig:Setup}b) become invisible to the probe, leading to the reduced absorption. These atoms gradually leave the near field and are replaced by incoming atoms in $|g\rangle$ both from far away and from ONF surface desorption, within a few nanoseconds, as being verified numerically (Fig.~\ref{fig:probe}e, the circles are from the numerical model detailed in Appendix~\ref{sec:mc}). This rapid recovery of the optical responses ensures that the aforementioned $10^7$ measurements within seconds are independent to each other.

\subsection{Modeling the ONF interface}\label{sec:model}

We numerically model the control and probe dynamics at the ONF interface (Fig.~\ref{fig:Setup}, Fig.~\ref{fig:probe}) both to predict the atomic state control efficiency $f({\bf r})$  (Eq.~(\ref{eq:f})) as those in Fig.~\ref{fig:Setup}d, and experimentally to infer the efficiencies $f({\bf r})$, $f_g({\bf r})$ (Eq.~(\ref{eq:fg})) from the $\overline{\delta T}$ measurements (Eq.~(\ref{eq:deltaT})). The model is based on optical Bloch equations (OBE)~\cite{scullyBook} (Appendix~\ref{sec:D1},~\ref{sec:D2}). Since the atomic state dynamics of interest is within the short nanosecond window, in this work we ignore the transition frequency shifts due to the surface interactions, which, averaged by the probe evanescent coupling, are at the MHz-level~\cite{Patterson2018,Solano2019}.

The numerical model starts with calculating the guided HE$_{11}$ field profiles ${\bf E}_c({\bf r})$, ${\bf E}_{\rm p}({\bf r})$ for the control and probe beams respectively~\cite{Tong2004} (Appendix~\ref{sec:circular}).  Next, for atom at location ${\bf r}$ in the near field, we integrate the Schr\"odinger equation for the resonant dipole D1-line interaction (Appendix~\ref{sec:D1}) to numerically obtain the evolution operator $U_c({\bf r},\{\mathcal{E}_j,\varphi_j\})$ for the picosecond composite pulses of interest. Since $v_{\rm T}\tau_c\ll \xi$, $\Gamma_a \tau_c\ll 1$, atomic motion and spontaneous decay can be ignored during the picosecond control.
With in mind the guided HE$_{11}$ profiles are invariant along $z$, we sample the control field in the $x-y$ plane (Fig.~\ref{fig:Setup}c) at a fixed $z$. We then evaluate the $|g\rangle-|a\rangle$ inversion efficiency $f({\bf r})$ (Eq.~(\ref{eq:fg}), Eq.~(\ref{eq:ga})) as those in Fig.~\ref{fig:Setup}d for atoms populating $|g\rangle$ with initial $\rho_{gg}^{(0)}=1$. On the other hand, for evaluating the ground-state depletion efficiency $f_g({\bf r})$ (Eq.~(\ref{eq:fg})), which is directly linked to the $\overline{\delta T}$ measurements, we set $\rho_{gg}^{(0)}=7/12$ and $\rho_{dd}^{(0)}=5/12$ for the $^{85}$Rb atom evenly populating the twelve ground-state Zeeman sublevels $|g_m\rangle, |d_m\rangle$ according to the experimental expectations. Here $m=m_F$ is the magnetic quantum number, see Fig.~\ref{fig:coupling} in the Appendix.

In contrast to the picosecond ``impulse'' interaction, one cannot simply ignore the atomic motion during the nanosecond D2 probe. In fact, with $v_{\rm T}\tau_{\rm p} \approx 380~{\rm nm}>\xi,\xi_{\rm p}$, we expect the thermal vapor-ONF system to be within a mesoscopic regime~\cite{Peyrot2019}, invalidating macroscopic effective media theory  based on local optical responses.
We set up an ``exact'' model and a diffusive average model 
to describe the D2 optical response at the ONF interface. 

For the ``exact'' model, we sample the phase space $({\bf r}_0,{\bf v}_0)$ of the thermal vapor uniformly surrounding the ONF, and evaluate the dipole interaction between the guided ${\bf E}_{\rm p}({\bf r}(t),t)$ and the atoms following the ballistic trajectories ${\bf r}(t)={\bf r}_0+{\bf v}_0t$. As mentioned, initially $\rho_0(0)=\rho^{(0)}$ uniformly populates all the 12 ground state Zeeman sublevels, which, after being subjected to the picosecond $U_c$ control, is denoted as $\rho(\tau_c)=U_c \rho_0(0) U_c^{\dagger}$. Then, during $\tau_c<t<\tau_c+\Delta t+\tau_p$, the $\rho(t)$ for the moving atoms, as well as $\rho_0(t)$ for those without experiencing the control pulses, all evolve according to the OBE on the D1-D2 manifold. To evaluate the probe scattering rates, we ignore the moderate ONF-modification to the atomic response~\cite{Solano2017,Solano2019} to simply have 
\begin{equation}
    \begin{aligned}
\bar \gamma(t)& =\frac{1}{\hbar} \left  \langle {\rm Im} [\langle{\bf E}^*_{\rm p}({\bf r(t)},t)\cdot {\bf d}\rangle_{\rho(t)}]\right \rangle_{{\bf r}_0,{\bf v}_0},\\
\bar \gamma_0(t)& = \frac{1}{\hbar} \left  \langle {\rm Im} [\langle{\bf E}^*_{\rm p}({\bf r(t)},t)\cdot {\bf d}\rangle_{\rho_0(t)}]\right \rangle_{{\bf r}_0,{\bf v}_0},
    \end{aligned}\label{eq:gamma1}
\end{equation}
with and without the control pulses. Here the inner $\langle...\rangle_{\rho(t)}$ represents ${\rm tr}(\rho(t)...)$ for the quantum mechanical average of a single trajectory. The outer $\langle ... \rangle_{{\bf r}_0,{\bf v}_0}$ sums the Monte Carlo trajectories according to the thermal distribution. The probe absorption $A$ without the control pulses, as well as the transient transmission $\overline{\delta T}$ induced by the control, are evaluated as
\begin{equation}
    \begin{aligned}
    A&=\frac{\hbar\omega_{\rm p} \int   \bar \gamma_0(t) {\rm d}t}{\int P_{\rm p}(t) {\rm d}t},\\
    \overline{\delta T} &=\frac{\int  (\bar \gamma_0(t)- \bar\gamma(t)){\rm d}t}{\int \bar \gamma_0(t) {\rm d}t },
    \end{aligned}\label{eq:deltaTt}
\end{equation}
in the limit of weak absorption $A\ll 1$. Here $P_{\rm p}(t)=\frac{1}{2}\int \varepsilon_0 \mathcal{N}^2 v_{\rm p} |{{\bf E}_{\rm p}({\bf r},t)}|^2 {\rm d}^2 {\bf r}_{\perp}$ is the transient optical power of the probe pulse, $v_{\rm p}=\omega_{\rm p}/\beta_{\rm p}$ is the phase velocity of the ONF guided probe light ($\beta_{\rm p}$ is the associated propagation constant), and $\mathcal{N}({\bf r}_{\perp})$ is the transverse ONF refractive index profile.

We note that with the external trajectories ${\bf r}(t)={\bf r}_0+{\bf v}_0 t$ at the 100~m/s level speed, mechanical forces associated with the surface interaction and the light pulses can be safely ignored. For trajectories that hit the ONF surface during the simulation time, we assume immediate desorption with a randomized emission direction, with the internal state reset to a random ground state. The simulations typically require $10^6$ sampling trajectories to converge for a specific control pulse configuration. We refer readers to Appendix~\ref{sec:mc} for details of the Monte Carlo simulations. 


The Monte Carlo simulation of the ``exact model'' becomes too resource-demanding for us when trying to explore the composite control configurations across the $\{\mathcal{E}_j,\varphi_j\}$ parameter space. To enhance the speed, a simpler diffusive average model is proposed for efficient evaluation of  $\overline{\delta T}$ as
\begin{equation}
    \begin{aligned}
\overline{\delta T}&\approx \overline{f_g({\bf r})}\\
&=\frac{\int R({\bf r}_{\perp}-{\bf r}_{\perp,0})i_{\rm p}({\bf r}_{\perp,0}) f_g({\bf r}_{\perp,0}) {\rm d}^2{\bf r}_{\perp} }{\int R({\bf r}_{\perp}-{\bf r}_{\perp,0})i_{\rm p}({\bf r}_{\perp,0}) {\rm d}^2{\bf r}_{\perp}},
    \end{aligned}\label{eq:dT2}
\end{equation}
using the ground state depletion ratio $f_g({\bf r})$ defined by Eq.~(\ref{eq:fg}). Here $i_{\rm p}({\bf r}_{\perp})\propto |{\bf E}_{\rm p}({\bf r}_{\perp})|^2$ is the transverse intensity profile of the guided D2 probe in the near field (Fig.~\ref{fig:probe}d). The thermal diffusion kernel is set as
\begin{equation}R({\bf r}_{\perp})= \frac{1}{\pi \xi_{\rm p}'^2}  e^{-|{\bf r}_{\perp}|^2/\xi_{\rm p}'^2}
\end{equation}
with a phenomenological $\xi'_{\rm p}\approx \sqrt{2} v_{\rm T}\tau_{\rm p}/2$ to account for atomic diffusion in the $x-y$ plane during the $\tau_{\rm p}$ probe time (see Appendix~\ref{sec:conv}). To account for the presence of ONF wall, we simply set $i_{\rm p}=0$ inside ONF.

Physically, Equation~(\ref{eq:dT2}) assumes that for evaluating the $\overline{\delta T}$ ratio in Eq.~(\ref{eq:deltaTt}), the coupling strengths between the guided probe and the surrounding atoms is decided by the local probe intensity $i_{\rm p}({\bf r})$ (Fig.~\ref{fig:probe}d) and the ground state population $\rho_{gg}$ only. The picosecond controls hardly modify the velocity distribution of the mesoscopic vapor. Therefore, while the aforementioned Doppler and transient broadening weaken the atom-light coupling strengths on average, the reductions are largely shared by the mesoscopic vapor with and without the control pulses.
The ignorance of coherence transients, as detailed in Appendix~\ref{sec:mc}, is justified by the fact that the effects are either averaged out during the $\tau_{\rm p}$ integration or largely canceled in the  $\overline{\delta T}$ ratio too.
In Appendix~\ref{sec:conv} we show that the Eq.~(\ref{eq:dT2}) model, with all the quite strong approximations,  generates results that agree fairly well with the ``exact model'' by Eq.~(\ref{eq:deltaTt}). The Eq.~(\ref{eq:dT2}) approximation helps us to efficiently simulate $\overline{\delta T}$ with the numerical model to compare with the experimental measurements, with which the actual $U_c({\bf r}, \{\mathcal{E}_j,\varphi_j\})$ parameters and the $f_g({\bf r})$, $f({\bf r})$ efficiencies in the experiments are inferred.

\subsection{Saturation of picosecond excitation}\label{sec:sat}



    \begin{figure}
        \centering
        \includegraphics[width=0.48\textwidth]{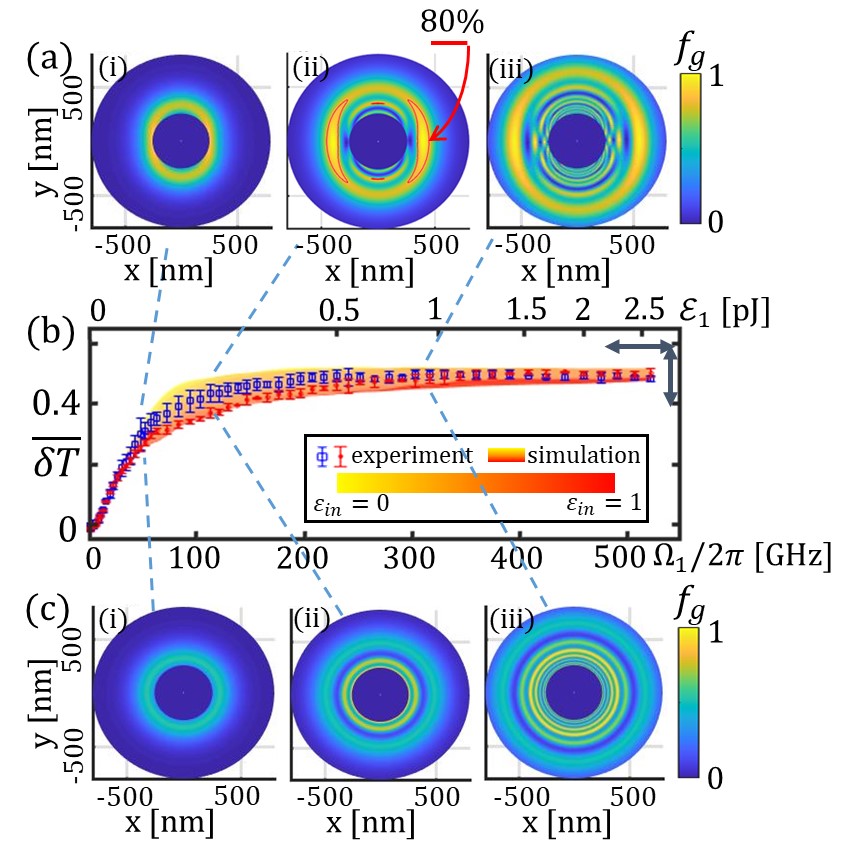}
        \caption{Polarization-dependent  picosecond D1 single excitation at the ONF interface. Figs.~(a)(c) give the simulated transient population depletion $f_g({\bf r})$ by a single pulse excitation with energy $\mathcal{E}_1$ (Eqs.~(\ref{eq:fg})~(\ref{eq:rhougg})). The incident HE$_{11}$ mode is with linear ($\varepsilon_{\rm in}=0$) and circular ($\varepsilon_{\rm in}=1$) incident ellipticity for the Figs.~(a)(c) simulations respectively.  A $f_g=80\%$ contour is highlighted in Fig.~(a,ii). Simulated $\overline{\delta T}$ according to Eq.~(\ref{eq:dT2}) at various incident ellipticities are given by Fig.~(b) with the rainbow plot, color-coded by $\varepsilon_{\rm in}$. The $\Omega_1$-axis is specified according to Eqs.~(\ref{eq:rRbi})(\ref{eq:Energy1}). The experimentally measured $\overline{\delta T}$ as a function of $\mathcal{E}_1$ and associated peak $\Omega_1$ are scatter-plotted with blue square ($\varepsilon_{\rm in}=0$) and red disk ($\varepsilon_{\rm in}=1$) symbols, with error-bars estimated from three repeated measurements. To obtain the theoretical-experimental match in Fig.~(b), the full experimental data set is uniformly rescaled (Appendix~\ref{sec:calib}) within the measurement uncertainties suggested by the double-sided arrows on the top right.}   
        \label{fig:OnePulse}
    \end{figure}

We first benchmark the ONF interface with an $N=1$, single-pulse control-probe experiment. Enhanced atom-light interactions have been demonstrated previously at the nanofiber interface~\cite{Spillane2008,Hendrickson2010,Venkataraman2011,Finkelstein2021}.  Here we demonstrate a full saturation of the control-induced change of probe transmission to $\overline{\delta T}\approx 50\%$ with merely sub-pico-Joule single control pulses. We highlight the polarization $\varepsilon_{\rm in}$-dependence of the transient optical response.


Intuitively, in presence of the inhomogeneous light intensity and ellipticity distributions (Fig.~\ref{fig:Setup}c, Fig.~\ref{fig:Circular}), we expect the impulse D1 excitation with large enough $\mathcal{E}_1$ to deplete the ground state population in a spatial dependent matter, lead to $f_g({\bf r})$ oscillating between 0 and 1 in the near field. Consequently, the evanescent coupling induced probe attenuation is expected to be transiently halved, $\overline{\delta T}\approx 50\%$~\cite{foot:cs}, as in Fig.~\ref{fig:probe}e at $\Delta t=0$.


In this section, the simple picture of optical saturation is confirmed by detailed measurements of $\mathcal{E}_1$-dependent $\overline{\delta T}$. To optimally retrieve the nonlinear signal, the probe delay is reduced to $\Delta t=0$~ps (Fig.~\ref{fig:probe}c). Typical transient transmission $\overline{\delta T}$ data are plotted in Fig.~\ref{fig:OnePulse}b as a function of pulse energy $\mathcal{E}_1$ and the peak Rabi frequency $\Omega_1$ estimated at the ONF surface (Appendix~\ref{sec:calib}).  In Appendix~\ref{sec:pol} we detail the HE$_{11}$ polarization control with automated polarization adjustments (Fig.~\ref{fig:probe}a). For the $\overline{\delta T}$ data here, the polarization states of control and probe beams are set to be linear ($\varepsilon_{\rm in}=0$, blue square symbols) and circular ($\varepsilon_{\rm in}=1$, red disk symbols) respectively. In both cases we find $\overline{\delta T}$ increases linearly with small $\mathcal{E}_1$. Furthermore,  nearly complete saturation of $\overline{\delta T}$ to $50\%$ occurs at $\mathcal{E}_1$ as small as 1~pJ. 






Following Sec.~\ref{sec:model}, we numerically evaluate the D1 atomic state dynamics subjected to the vectorial light-atom interaction in the near field. Typical ground-state depletion efficiency $f_g({\bf r})$ are plotted in Fig.~\ref{fig:OnePulse}(a)(c) for the case of linear ($\varepsilon_{\rm in}=0$) and circular ($\varepsilon_{\rm in}=1$) HE$_{11}$ modes of control respectively. Notice for the $N=1$ pulse here, $\tau_c\ll 1/\omega_{{\rm hfs},g},1/\omega_{{\rm hfs},g}$ is fairly well satisfied, so that the $J_g=1/2\leftrightarrow J_a=1/2$ intuition in Appendix~\ref{sec:composite} can be applied. In particular, as in Fig.~\ref{fig:OnePulse}a, the ground state population can be nullified in the near field where the polarization is purely linear. For comparison, the reduced peak $f_g$ for the circular incident polarization in Fig.~\ref{fig:OnePulse}c is associated with local ellipticity $\varepsilon\approx 0.95$ (Appendix~\ref{sec:circular}) so the $\sigma^\pm$ coupling strengths are different substantially (Eq.~(\ref{eq:omJI})), making perfect $|g\rangle-|a\rangle$ inversion impossible in presence of population in different Zeeman sublevels. Nevertheless,  regardless of the $\varepsilon_{\rm in}$ value, the near-field chirality~\cite{Lodahl2017} prevents completely circular polarization (Fig.~\ref{fig:composite}), {\it i.e.} $\varepsilon(x,y)=1$, from uniformly occurring in the near field. In this case, as illustrated by Fig.~\ref{fig:composite}, all ground state sublevels have chance to be strongly excited by the control pulse, leading to $\varepsilon_{\rm in}$-independent $\overline{\delta T}$ saturation to $\sim50\%$ at large enough $\mathcal{E}_1$ in Fig.~\ref{fig:OnePulse}b.

We present simulated $\overline{\delta T}$ according to Eq.~(\ref{eq:dT2}) in the same Fig.~\ref{fig:OnePulse}b as color domain plot, with various ellipticity color-coded for the incident control (and the orthogonal probe) polarizations. The experimentally measured data are matched to the simulation, by linearly rescaling the $\overline{\delta T}$, $\mathcal{E}_1$ and $\Omega_1$ axes within the uncertainties by the ``steady state'' absorption $A$ and laser power measurements (Appendix~\ref{sec:calib}). Fairly good agreements are found between theoretical and experimental $\overline{\delta T}$. The notable discrepancies near $\Omega_1=2\pi\times 100~$GHz could be due to imperfect control/probe polarization control in the experiment (Appendix~\ref{sec:pol}).



\subsection{Robust composite control at the ONF interface}\label{sec:three}

\begin{figure*}
    \centering
    \includegraphics[width=1\textwidth]{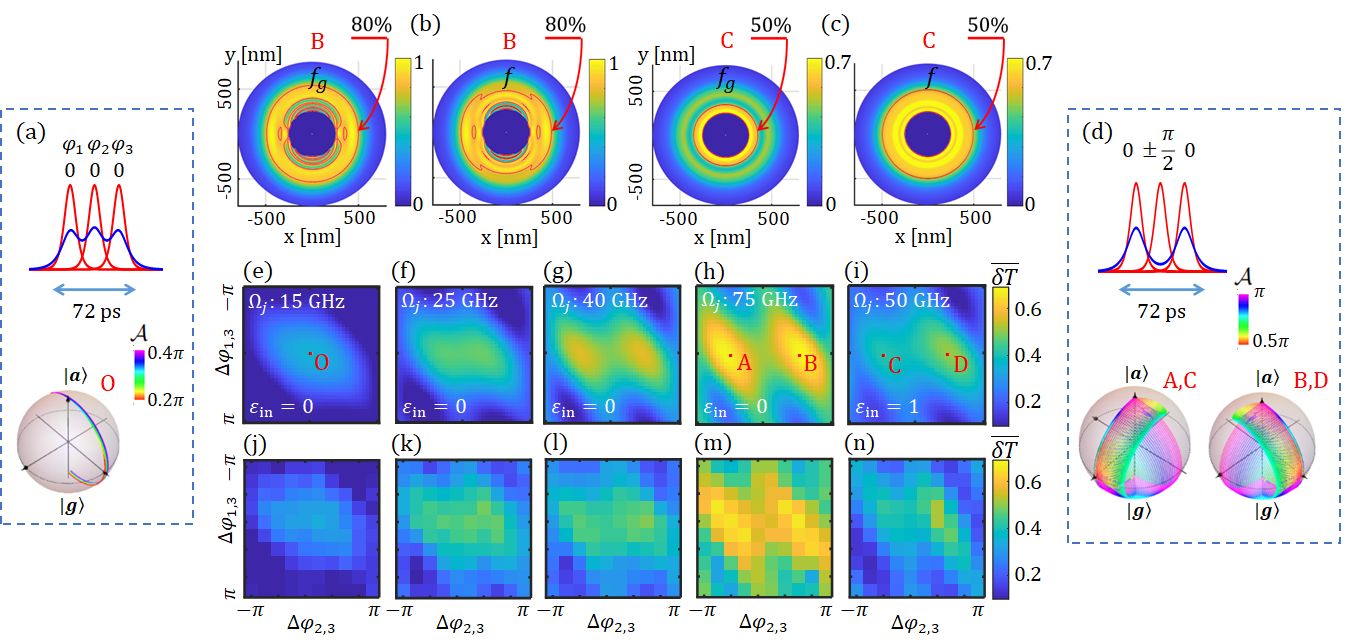}
    \caption{Three-pulse atomic state control through the ONF interface. (a) and (d): The top plots give the intensity $|{\bf E}_c|^2$  (red) and the real envelope ${\rm Re}[E_c]$  (blue) of the $N=3$ pulse sequences shaped from a transform-limited $\tau_0=12$~ps pulse, with $\tau_{\rm d}$=24~ps inter-pulse delay, and $\{\varphi_j\}=0$ and $\{\varphi_j\}=\{0,\pm\pi/2,0\}$ respectively. The bottom plots give the associated Bloch sphere 2-level dynamics with a broadened sub-pulse area $\mathcal{A}_j=\mathcal{A}$. (b)(c): Simulated $f_g({\bf r})$ and $f({\bf r})$ for the Fig.~(d) $\{\varphi_j\}$ combinations. The red capital letters in Figs.~(a-d) mark the parameter combinations in Figs.~(e-i). (e-i): Simulated transient transmission  $\overline{\delta T}$ vs $\Delta\varphi_{1,3}$ at various peak $\Omega_j$, according to Eq.~(\ref{eq:rRbi}) with the associated  $\mathcal{E}_j=\mathcal{E}=2, 6, 17,50,20\times 10^{-3}$~pJ sub-pulse energy (Eq.~(\ref{eq:Energy1})). (j-n): Corresponding experimental $\overline{\delta T}$ data. The experimental $\Omega_j$, $\overline{\delta T}$ values are globally re-scaled to match the simulations, detailed in Appendix~\ref{sec:calib}. }
    \label{fig:ThreePulse}
\end{figure*}


We now demonstrate robust population inversion at the ONF interface. This is achieved by implementing the composite technique prescribed by ref.~\cite{Genov2014} with our picosecond pulse sequence generator (Fig.~\ref{fig:generation}). Due to technical reasons to be discussed in Sec.~\ref{sec:largeN}, we limit the sub-pulse number to $N=3$ in this demonstration. 

As outlined in Appendix~\ref{sec:generator}, the picosecond sequence is generated by shaping $\tau_0=12$~ps pulses into three sub-pulses, with $\tau_{\rm d}=24$~ps inter-pulse equal delay, equal pulse energy $\{\mathcal{E}_j\}$,  and independently programmable phases $\{\varphi_j\}$. At a fixed $\mathcal{E}_j=\mathcal{E}$, we perform 2D scan of the relative phases $\Delta \varphi_{1,3}=\varphi_1-\varphi_3$ and $\Delta \varphi_{2,3}=\varphi_2-\varphi_3$ in small steps to record the transient transmission $\overline{\delta T}$ (Fig.~\ref{fig:ThreePulse}(j-n)). 
As in Figs.~\ref{fig:ThreePulse}(a)(d), 
the moderate $\tau_{\rm d}=2\tau_0$ leads to partially overlapping sub-pulses. With accurate modeling in Appendix~\ref{sec:D1} to account for the overlap, the composite scheme is effectively captured by  non-overlapping pulses, as assumed in the following. 


At low pulse energy, we expect the picosecond control to be most efficient when $\Delta \varphi_{1,3}=\Delta \varphi_{2,3}=0$ so the resonant spectra density to $|g\rangle-|a\rangle$ excitation is maximized. This is the case in Fig.~\ref{fig:ThreePulse}e according to the simulation where the equal-phase point with optimal $\overline{\delta T}$ is marked with ``O''. The corresponding optical waveform is plotted in Fig.~\ref{fig:ThreePulse}a on the top. Indeed, for the equal-phase case, the 3-pulse control is equivalent to a single-pulse control with an elongated duration $\tau_c$. Therefore, the control dynamics is not robust against variation of optical intensity (nor polarization), as suggested by the simplified 2-level Bloch sphere dynamics in the bottom plot of Fig.~\ref{fig:ThreePulse}a.

However, with increased single-pulse energy $\{\mathcal{E}_j\}$ and the associated $\{\Omega_j\}$, a transition of the optimal $\{\varphi_j\}$ occurs around $\mathcal{A}_{j}\approx \pi/3$. Beyond the point, the optimal $\varphi_2$ takes none-zero values relative to the equal $\varphi_{1,3}$. This is suggested by the simulations in Figs.~\ref{fig:ThreePulse}(e-h), which agree globally with the Figs.~\ref{fig:ThreePulse}(j-m) experimental measurements without freely adjustable parameters (Sec.~\ref{sec:calib}). The subtle difference between Fig.~\ref{fig:ThreePulse}(e-i) and Fig.~\ref{fig:ThreePulse}(j-n) are likely associated with slightly varying $\mathcal{E}_j$ during the experimental phase scan (Sec.~\ref{sec:largeN}). With $\Omega_j \approx 2\pi \times 75$~GHz and linear $\varepsilon_{\rm in}=0$ (Fig.~\ref{fig:ThreePulse}(h,m)), transient transmission $\overline{\delta T}\sim 70\%$ are found at two phase combinations with $\varphi_2=\pm \pi/2, \varphi_{1,3}=0$. For the case of $\varphi_2=\pi/2$, our full level simulations suggest $\overline{\delta T}$ integrated over $\tau_p=2$~ns is supported by transient depletion of the $|g\rangle$ state population with $f_g ({\bf r})>80\%$ (Fig.~\ref{fig:ThreePulse}b), over a connected area at the ONF proximity by the picosecond impulse control to be substantially larger than the 1-pulse case (Fig.~\ref{fig:OnePulse}a). This population depletion is largely due to the $|g\rangle-|a\rangle$ inversion with $f({\bf r})>80\%$ albeit across a slightly smaller area.


The efficient population inversion across the highly inhomogenuous near field, as suggested by the Fig.~\ref{fig:ThreePulse}(h,m) data, is a result of geometric robustness inherent to the  2-level composite control~\cite{ICHIKAWA2012} (also see Appendix~\ref{sec:composite}).  As illustrated in Fig.~\ref{fig:ThreePulse}d, the ``redundant'' SU(2) rotation enables robust $|g\rangle-|a\rangle$ inversion for the equal-area 3-pulse control with $\mathcal{A}_j=\mathcal{A}\in(0.5\pi,\pi)$ when $\varphi_2$ takes the value of $\pm \pi/2$ relative to $\varphi_{1,3}$. In both cases, the rotation by the 2nd sub-pulse automatically cancels out the extra rotations by the first and the third sub-pulses. As a result, an $f>80\%$ inversion efficiency is uniformly achieved in the near field (Fig.~\ref{fig:ThreePulse}b) to support the $\overline{\delta T}\approx 70\%$ observation (Fig.~\ref{fig:ThreePulse}m) for the thermal atoms. We also refer readers to Fig.~\ref{fig:Setup}(d,ii)(e) for the $N=5$ example~\cite{Genov2014}, where the full-level simulation suggests nearly perfect inversion over a substantial volume in the near field.


We finally remark on Raman transitions driven by the composite pulse with the fairly long $\tau_c\sim 1/\omega_{{\rm hfs},g}$ duration (Appendix~\ref{sec:JI}). 
As in Fig.~\ref{fig:ThreePulse}b, compared to the expected inversion efficiency $f({\bf r})$, the  $\rho_{gg}$-depletion efficiency $f_g({\bf r})$ is larger, which is a result of directional $|g\rangle\rightarrow|d\rangle$ transfer at $\varphi_2=\pi/2$. Similarly, not shown in Fig.~\ref{fig:ThreePulse}b is the ``A'' point with $\varphi_2=-\pi/2$, where an opposite $|d\rangle\rightarrow|g\rangle$ transfer leads to reduced $f_g({\bf r})$ relative to $f({\bf r})$. However, in neither case these Raman contributions notably affect the apparent $\pm \pi/2$ symmetry in the transmission $\overline{\delta T}$ in Figs.~\ref{fig:ThreePulse}(e-h) and Figs.~\ref{fig:ThreePulse}(j-m), since when the local polarization is approximately linear  ($\varepsilon({\bf r})\ll 1$, Fig.~\ref{fig:Setup}c), the Raman transitions are largely suppressed as long as $\omega_{{\rm hfs},a} \tau_c\ll 1$~\cite{Happer1972} (Appendix~\ref{sec:JI}). On the other hand, in Fig.~\ref{fig:ThreePulse}(i)(n)  $\overline{\delta T}$ at $\varphi_2=-\pi/2$ (the ``C'' point) is substantially smaller than that for $\varphi_2=\pi/2$ (the ``D'' point). The broken sign symmetry is associated with substantial $|g\rangle \leftrightarrow |d\rangle$ Raman transfer, as unveiled by comparing $f_g({\bf r})$ with $f({\bf r})$ in Fig.~\ref{fig:ThreePulse}c according to the full-level simulations. Here, with the control light in the circularly polarized HE$_{11}^{\sigma}$ mode, the ellipticity in the near field is substantially larger (Appendix~\ref{sec:circular}). With atoms randomly initialized in $|g\rangle$ and $|d\rangle$, the apparently inefficient depletion $f_g({\bf r})$ at $\varphi_2=-\pi/2$ in Fig.~\ref{fig:ThreePulse}c is associated with substantial $|d\rangle \rightarrow |g\rangle$ transfer which negatively offsets the depletion by $|g\rangle-|a\rangle$ inversion. Similarly, not shown in Fig.~\ref{fig:ThreePulse} are $f_g({\bf r})>f({\bf r})$ at  ``D'' point with $\varphi_2=\pi/2$, as those in Fig.~\ref{fig:ThreePulse}b, but by a wider margin, due to the more efficient $|g\rangle \rightarrow |d\rangle$ transfer. We note that the picosecond Raman excitation is also expected to induce $|g\rangle-|d\rangle$ 2-photon coherence, resulting in $\omega_{{\rm hfs},g}$-frequency transients in the nanosecond probe transmission (Fig.~\ref{fig:probetransient}). These transients are efficiently averaged out with the $\tau_{\rm p}=2$~nanosecond integration (Fig.~\ref{fig:probe}(b,c)), not resolved by the $\overline{\delta T}$ measurements in this work.

\section{Discussions}\label{sec:discussion}

\subsection{Toward large $N$}\label{sec:largeN}

The $N=3$ composite picosecond control demonstrated in Sec.~\ref{sec:three} relies on an ergodic search for optimal pulse parameters. As in Fig.~\ref{fig:ThreePulse}, the method supports detailed investigation of the control dynamics across the parameter space by comparing the experimental measurements with theory. On the other hand, the ``brutal force'' approach becomes impractical at larger $N$, particularly when the search time is constrained by slow experimental cycles. Ideally, the relative amplitude and phase $\{a_j,\varphi_j\}$ should be directly programmed into an $N$-pulse sequence generator according to the optimal control theory for the accurately modeled experimental system. When the physical model of either the interaction or the pulse shaper itself is not accurate, then a close-loop approach should be followed for in situ optimizing of pulse parameters, similar to the pioneer works in nonlinear optics~\cite{Goswami2003,Wollenhaupt2011} and quantum information processing~\cite{Kelly2014}. 

Efforts toward composite control at larger $N$ in this work is frustrated by a pulse shaper parameter cross-talk, as mentioned in ref.~\cite{Ma2020}. As being discussed there, the cross-talk is associated with acousto-optical transduction, in particular the nonlinearity of multi-frequency rf amplification for driving the single AOM in this work  (Appendix~\ref{sec:generator}). The cross-talk leads to enough complexity to prevent us from precisely modeling the shaper itself when operating at the required efficiency for $N\geq 5$.  The cross-talk was also large enough to prevent a successful ``close-loop'' optimization with our ONF setup. Toward directly programming optimal picosecond control, we are currently working on improving the pulse shaper for efficient arbitrary sequence generation at large $N$~\cite{foot:shaper}.

\subsection{Summary and outlook}

A fundamental quest in nanophotonics is to enhance optical nonlinearities through confinements. The enhancements are not only instrumental to realizing efficient nonlinear optics and spectroscopy~\cite{Piatkowski2016, Dombi2020}, but also may support controllable interaction mediated by single confined photons~\cite{Chang2018}. In this work, we suggest that highly precise, arbitrary control of atomic electric dipole transitions can be achieved within picoseconds at nanophotonic interfaces, despite the coupling strength inhomogenuity, using the NMR-inspired composite technique in a power-efficient manner. It is important to note that this kind of atomic state control is rarely achieved before, even in free space~\cite{Ma2020}.


Experimentally, this work takes a first step toward precise nanophotonic control with the composite picosecond scheme. An optimally phased $N=3$ sequence is demonstrated to robustly invert the population of free-flying atoms across an optical nanofiber. In particular, the $\sim 70\%$ reduction of the evanescently coupled probe absorption, integrated over two nanoseconds of mesoscopic atomic motion~\cite{Peyrot2019}, strongly suggests $f>80\%$ population inversion {\it uniformly} achieved around the nanofiber in the near field. We confirm the accurate implementation of the geometric scheme~\cite{Genov2014,ICHIKAWA2012} by matching the measurements with first-principle modeling of the mesoscopic light-atom interaction. Our experimental work paves a practical pathway toward $N\geq5$ composite control~\cite{Genov2014, Low2016} of atomic state, for atoms confined in the near field~\cite{Vetsch2010,Meng2018,Su2019}, with exquisite precision.


Composite pulses are widely applied across fields to achieve robust control of meta-stable quantum states~\cite{Kabytayev2014, Zanon-Willette2018, Saywell2020a,Dreissen2022a, Qiu2022,Bluvstein2022}. We expect many novel applications by extending the technique to control two-level atoms at nanophotonic interfaces. For example, it is well known that the optical transition properties are modified at nanophotonic interfaces. By measuring the line centers and widths and to compare with the free-space values, the nanoscopic electro-magnetic perturbations including the van~der~Waals interaction can be inferred~\cite{Fuchs2018,Patterson2018,Solano2019c,Peyrot2019a,Hummer2021}. Here, instead of relying on regular linear spectroscopy, precise $\pi/2$ and $\pi$ pulses can be combined to optimize the measurement efficiency and to auto-balance the light shifts~\cite{Sanner2018,Yudin2020}.  
More generally, the highly precise optical dipole control should facilitate novel developments of integrated nonlinear optics~\cite{Spillane2008,Hendrickson2010,Venkataraman2011,Finkelstein2021} by improving the contrast of nonlinear optical modulation to the limit set by 2-level atoms. Repetitive application of the fast, arbitrary controls might even enable one to dynamical decouple or amplify time-dependent perturbations on demand, similar to those envisioned in NMR spectroscopy~\cite{Yuge2011,Wang2022e}.

Finally, we note the composite picosecond control can be applied to trapped array of atoms~\cite{Vetsch2010,Meng2018, Su2019} for efficient steering of collective couplings of the atomic ensemble to the waveguide~\cite{Scully2015,He2020a,He2020b}. To illustrate the effect, we come back to the $N=5$ scheme in Fig.~\ref{fig:Setup} and consider a 1D lattice gas trapped near ONF~\cite{Vetsch2010,Meng2018,Su2019}. Following the $|g\rangle-|a\rangle$ inversion by the $N=5$  composite pulse~\cite{Genov2014},  a second $N=5$ composite pulse can drive a nearly perfect return of atomic population back to the ground states. If the second composite pulse is sent from the opposite direction of the ONF guide, then a sub-wavelength-scale $e^{2 i \beta_{\rm c} z}$ optical phase is patterned to the state $|g\rangle$. Here $\beta_{\rm c}$ is the propagation constant of the guided control pulses. As such, the delocalized $|g\rangle-|e\rangle$ dipole spin wave excited beforehand by the guided ${\bf E}_{\rm p}$ pulse to the lattice~~\cite{Vetsch2010,Meng2018, Su2019} can be reversibly shifted into the subradiant domain on demand~\cite{Scully2015,He2020a, He2020b}, for accessing the nearly dissipationless, long-range optical dipolar interaction dynamics at the ONF interface~\cite{Albrecht2017a,
    Noh2017, Chang2018, Kornovan2019, Zhang2019, Zhang2020b,Rui2020, Buonaiuto2021,He2021a}.




\section*{Acknowledgement}
We thank Professor Darrick Chang for very helpful discussions. We acknowledge support from National Key Research Program of China under Grant No.~2022YFA1404204 and No.~2017YFA0304204, from National Natural Science Foundation of China under Grant No.~12074083, 61875110, 62105191, 62035013, 62075192.

\section*{Data availability}
Data and simulation codes underlying the results presented in this paper are not publicly available at this time but may be obtained from the authors upon reasonable request.

\appendix

\section{Composite control of an $I=0$ alkaline atom}\label{sec:composite}

\begin{figure}
        \centering
        \includegraphics[width=0.45\textwidth]{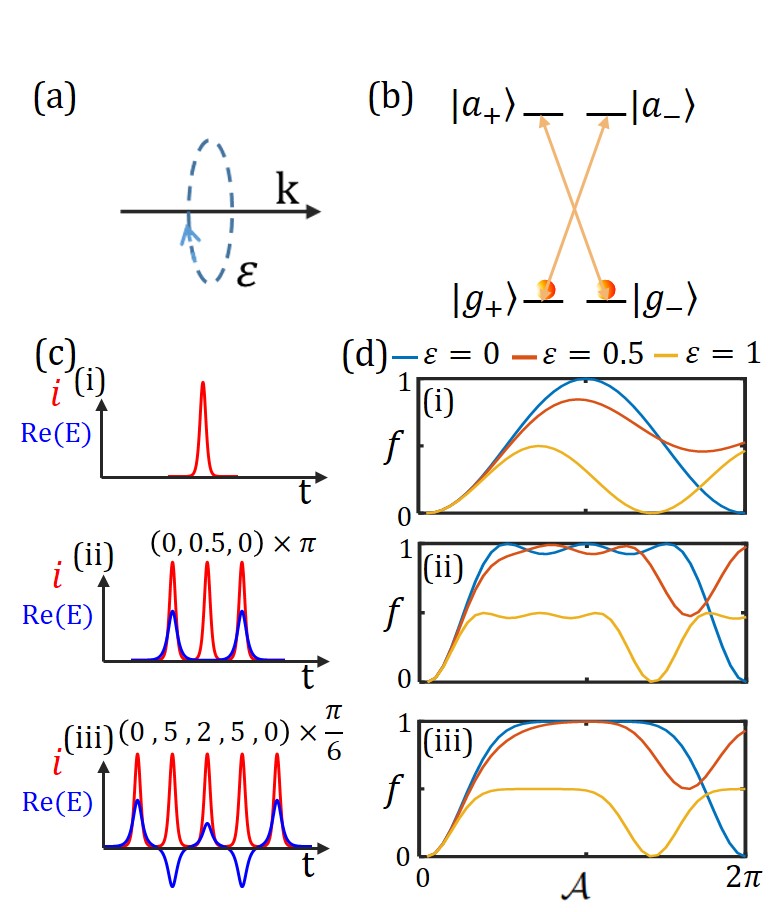}
        \caption{D1 population inversion by resonant composite excitations, for an alkaline atom without hyperfine structure.  The level diagram for the degenerate $S_{1/2}-P_{1/2}$ transition is given in (b).  The quantization axis is set along the light helicity vector ${\bf e}_h = {\bf E}_c^*\times {\bf E}_c/|{\bf E}_c|^2$, which is parallel to ${\bf k}$ for the free field example. The ellipticity is  defined as $\varepsilon=|{\bf e}_h|$.
     Fig.~(d) gives numerical results of inversion efficiency $f$ at various $\varepsilon$ for $N=$1, 3, and 5 pulses, as a function of single  pulse area $\mathcal{A}=|\int  \Omega_c {\rm d}t|$, according to Eq.~(\ref{eq:rRbi}). The Fig.~(c) plots on the left give the time-dependent intensity $i$ (red line) and real envelope ${\rm Re}[E_c]$ (blue line) of the pulses, with relative phases according to ref.~\cite{Genov2014} marked on the top. } 
    \label{fig:composite}
\end{figure}

In this Appendix, we show that for alkaline atoms and if the hyperfine splitting can be ignored, then the composite picosecond control can operate well within a fine-structure manifold. We consider the exemplary composite technique to invert the state population of a fictitious alkaline atom without hyperfine structure ($I=0$), by driving the $J_g=1/2\rightarrow J_a=1/2$ D1 line with free-space optical pulses. The conclusions are applicable to the D2 line, and to more general SU(2) controls~\cite{Low2016}  in a straightforward manner. The simple picture is  also applicable locally in the near field (Appendix~\ref{sec:circular}), and to $I\neq0$ atom in the $\tau_c\ll 1/\omega_{{\rm hfs},g},1/\omega_{{\rm hfs},a}$ limit (Appendix~\ref{sec:JI}). 

As in Fig.~\ref{fig:composite}, in free space the quantization axis for the light-atom interaction is naturally chosen along the ${\bf k}$ direction. More generally, for the incident control electric field characterized by a slowly varying complex envelope ${\bf E}_c$ and a helicity vector,
\begin{equation}
{\bf e}_h={\bf E}_c^*\times {\bf E}_c/|{\bf E}_c|^2,\label{eq:eh}
\end{equation}
the D1 transition is decomposed into $\sigma^+$ and $\sigma^-$ transitions along the ${\bf e}_h$ quantization axis, with the associated Rabi frequencies
\begin{equation}
    \begin{array}{l}
\tilde{\Omega}_{\rm c}^+({\bf r},t)=\sqrt{2}{\rm cos}(\theta/2)\Omega_{\rm c}({\bf r},t)\\
\tilde{\Omega}_{\rm c}^-({\bf r},t)=\sqrt{2}{\rm sin}(\theta/2)\Omega_{\rm c}({\bf r},t).
    \end{array}\label{eq:omJI}
\end{equation}
Here $\theta\in [0,\pi/2] $ is associated with the field ellipticity 
\begin{equation}\varepsilon=|{\bf e}_h|
\end{equation} 
as $\varepsilon={\rm cos}\theta$. The Rabi frequency $\Omega_c$ is defined as
\begin{equation}
    \Omega_{c}({\bf r},t)= \frac{E_{\rm c}({\bf r},t)}{\hbar}\frac{1}{\sqrt{3}}|\langle J_g||{\bf d}||J_a\rangle|.\label{eq:rRbi}
\end{equation}
It is important to note that to apply Eqs.~(\ref{eq:eh})(\ref{eq:omJI}) for the linearly polarized ${\bf E}_c$, the quantization axis needs to be perpendicular to ${\bf E}_c$, {\it i.e.}, as a limiting case of small $\varepsilon$.  

Here we consider $\Omega_{\rm c}({\bf r},t)=\Omega_{\rm c}(t)$ and ${\bf e}_h({\bf r})={\bf e}_h$ to be spatially uniform. Our goal is to design certain pulse sequences to invert the population of the reduced D1 system initialized in the unpolarized ground state, $\rho(0)=\frac{1}{2}(|g_+\rangle\langle g_+|+|g_-\rangle\langle g_-|)$. This is investigated 
with the simulation method outlined in Appendix~\ref{sec:D1} by solving the Schr\"odinger equation for a control time $\tau_c$ and then evaluate $f=\rho_{a a}\equiv \rho_{a_+ a_+}+\rho_{a_- a_-}$, for the $J_g=1/2\leftrightarrow J_a=1/2$ transition.

Clearly, the Rabi frequencies for the $\sigma^{\pm}$ couplings by Eq.~(\ref{eq:omJI}) are equal only for linearly polarized light. For general polarization state, it is not possible to simultaneously invert the two sub-spins with $N=1$ single ``$\pi$''-pulse (Fig.~\ref{fig:composite}(c,i)). In fact, in the limiting case of circular polarized $\pi$-pulse (Fig.~\ref{fig:composite}(d,i), the yellow curve with $\varepsilon=1$), only $50\%$ of ground-state population can be inverted, albeit with a $\sqrt{2}$-times larger Rabi oscillation frequency relative to the linear polarized case (Fig.~\ref{fig:composite}(d,i), the blue curve with $\varepsilon=0$). For the linear polarization, the inversion near $\mathcal{A}=\pi$ scales as $\rho_{aa}(\tau_c)={\rm sin}^2(\mathcal{A}/2)$ and is therefore quite sensitive to the pulse area $\mathcal{A}=|\int\Omega_c {\rm d}t|$, requiring perfect control of light intensity. 

For comparison, in Fig.~\ref{fig:composite}(d,ii) and Fig.~\ref{fig:composite}(d,iii) the population inversion efficiencies are shown for a composite 3-pulse (Fig.~\ref{fig:composite}(c,ii)) and 5-pulse (Fig.~\ref{fig:composite}(c,iii)) sequence respectively. Both sequences follow prescription by ref.~\cite{Genov2014} as $\mathcal{A}_j=\mathcal{A}$ close to $\pi$ and phase $\{\varphi_j\}$=$\{0,\pi/2,0\}$, $\{0,5\pi/6,\pi/3,5\pi/6,0\}$ for the 3- and 5-pulses. The 3-pulse sequence is experimentally exploited in Sec.~\ref{sec:three}. The geometric origin of the intensity-error resilience for the $N=3,5$ composite pulses are illustrated with the Bloch sphere picture in Fig.~\ref{fig:ThreePulse}d, Fig.~\ref{fig:Setup}e respectively. Here, comparing with the single pulse inversion in Fig.~\ref{fig:composite}(d,i) that is perfected only for linearly polarized light at $\mathcal{A}=\pi$, the composite pulse schemes are much more tolerant to deviation of  $\mathcal{A}$ from $\pi$, and can achieve $f\rightarrow 1$ even by elliptically polarized excitations. 


\section{The HE$_{11}$ mode}\label{sec:circular}

\begin{figure}
        \centering
        \includegraphics[width=0.48\textwidth]{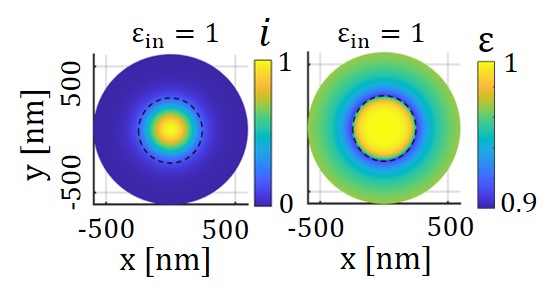}
        \caption{Near field distribution of the normalized light intensity $i(x,y)$ and ellipticity $\varepsilon(x,y)$ for the circular HE$_{11}^{+}$ mode of the ONF in this work at $\lambda_{\rm c}=795$~nm. } 
    \label{fig:Circular}
\end{figure}

Aided by precise knowledge of the nanofiber geometry~\cite{Su2019}, we follow ref.~\cite{Tong2004} to calculate the field distribution around the nanofiber for the guided HE$_{11}$ mode at specific wavelengths. The intensity and ellipticity distribution for the control laser at $\lambda_{\rm c}=795$~nm, for the case of linearly polarized HE$_{11}^y$ mode with  $\varepsilon_{\rm in}=0$ is presented in Fig.~\ref{fig:Setup}c. For the ${\bf e}_y$ polarized incidence, we see the near field ${\bf E}_{{\rm HE}_{11}^y}$ is linearly polarized around $y=0$ with $\varepsilon({\bf r})\ll 1$. On the other hand, for control pulses with circular polarized HE$_{11}^{+}$ mode with incident ellipticity $\varepsilon_{\rm in}=1$, the electric field ${\bf E}_{{\rm HE}_{11}^+}={\bf E}_{{\rm HE}_{11}^x}+i{\bf E}_{{\rm HE}_{11}^y}$ has a uniform local ellipticity $\varepsilon({\bf r})\approx 0.95$ which is substantially larger as shown here in Fig.~\ref{fig:Circular}. Similar mode intensity and polarization distributions are numerically evaluated for the probe pulse at $\lambda_{\rm p}=780$~nm.

More generally, for ${\rm HE}_{11}$ mode with an incident $\varepsilon_{\rm in}={\rm cos}(\Theta)$ ellipticity, with an angle $\phi$ between the incident elliptical axis and the $x$-axis, the field is a coherent superposition of that for the HE$_{11}^+$ and HE$_{11}^-$ modes weighted by ${\rm cos}(\Theta/2)$ and ${\rm sin}(\Theta/2)$, given by 
\begin{equation}
{\bf E}({\bf r})={\rm cos}(\frac{\Theta}{2}) e^{i\frac{\phi}{2}} {\bf E}_{{\rm HE}_{11}^+}+{\rm sin}(\frac{\Theta}{2}) e^{-i\frac{\phi}{2}} {\bf E}_{{\rm HE}_{11}^-}.\label{eq:hyb}
\end{equation}
The normalized intensity distribution $i({\bf r})$ and ellipticity distribution $\varepsilon({\bf r})$ can be evaluated accordingly. The spatial profile of the evanescent field is equipped to evaluate the light-atom interaction to be detailed in the following.


\section{Modeling the D1 control}\label{sec:D1}

\subsection{Level diagram}
The level diagram for the full D1/D2 electric dipole interaction is summarized in Fig.~\ref{fig:coupling}a. For the convenience of numerical calculation, we choose the ONF guiding direction  ${\bf e}_z$, along  which the field intensity and polarization distributions are invariant, as the fixed atomic quantization axis. The choice of local helicity axis as quantization axis will be discussed in Appendix~\ref{sec:JI}. To conveniently formulate the multi-level vectorial interactions, we introduce Dirac kets $|g_m\rangle$,$|d_m\rangle $ to label the Zeeman sublevels of the hyperfine ground states. Similarly, the excited state Zeeman sublevels are label by $|a_m\rangle$, $|e_m\rangle$. The $a,e$ symbols also index the total angular momentum $F$ of the corresponding hyperfine levels~\cite{Qiu2022}.


\subsection{The D1 Hamiltonian}
We consider spectrum transform-limited picosecond D1 pulses from a mode-locked laser with a temporal amplitude profile $\mathcal{P}(t)={\rm sech}(1.76 t/\tau_0)$~\cite{sech2pulse}. After time-domain pulse shaping~\cite{Ma2020}, the composite sequence is sent through ONF to interact with atom at location ${\bf r}$. The pulsed optical field at the ONF interface is described by a slowly-varying envelop function 
\begin{equation}
{\bf E}_{\rm c}({\bf r},t)={\bf E}({\bf r})S_N(\{a_j,\varphi_j\},t), \label{eq:Ec}
\end{equation} 
with a spatial profile according to Eq.~(\ref{eq:hyb}), and temporally following the composite profile according to Eq.~(\ref{equ:inOut}),
\begin{equation}
S_N(\{a_j,\varphi_j\},t)=\sum_{j=1}^{N} a_j e^{i\varphi_j} \mathcal{P}(t-(j-1)\tau_{\rm d}),\label{eq:cPulse}
\end{equation}
with $|a_j|\leq 1$ and $\tau_d\gg\tau_0$.

With the atomic states defined earlier, the Rabi frequencies to drive the $|g\rangle-|a\rangle$ and $|d\rangle-|a\rangle$ transitions are written as
\begin{equation}
\Omega_{a_n s_m}^{l}({\bf r},t)=\langle a_{n}|{\bf E}_{\rm c}({\bf r},t) \cdot {\bf d}_{l}|s_m\rangle/\hbar
\end{equation}
for all the $s=g,d$ states. Here ${\bf d}_l$ with $l=-1,0,1$ are the electric dipole operators of the atom along $\{{\bf e}_-,{\bf e}_z, {\bf e}_+\}$ directions respectively. Therefore, $n=m+l$ is required by conservation of the magnetic quantum number. More generally, with the Clebsch-Gordan coefficients $\mathcal{C}_{a_n c_m}^{l}$ and the D1 Rabi frequency defined in Eq.~(\ref{eq:rRbi}), the Rabi frequencies can also be written as $\Omega_{a_n s_m}^l({\bf r},t)=\sqrt{3}\Omega_{\rm c}({\bf r},t)\mathcal{C}_{a_n s_m}^l$.

The D1 resonant dipole interaction under the rotating wave approximation is written as
    \begin{equation}
\begin{array}{l}            
    H_{\rm D1}({\bf r},t) = \hbar\sum_{a}(\omega_{a}-\omega_{a 0})\sigma^{a_n a_n} + \\   
    ~~~~~~~~~~~~~~~~\hbar\sum_{s=g,d} (\omega_{s}-\omega_{g0})\sigma^{s_m s_m}+
    \\
    ~~~~~~~~~~~~~~~~\frac{\hbar}{2}\sum_{s={g,d}}\sum_l\Omega^{l}_{a_n s_m}({\bf r}, t)\sigma^{a_n s_m} + {\rm h.c.}
        \end{array}\label{eq:HD1}
    \end{equation}
Here an implicit summation of repeated $n,m$ indices are assumed. The $\omega_{a0}, \omega_{g0}$ are decided by the energy of reference level in the 5P$_{1/2}$ and 5S$_{1/2}$ electronic states respectively, chosen as the top hyperfine levels in this work. The control Rabi frequencies are accordingly written in the $\omega_{a0,g0}$ frame with the resonant frequency $\omega_c=\omega_{a0,g0}$ canceled. The Pauli matrices are defined as $\sigma^{a_n s_m}=|a_n\rangle\langle s_m|$, similarly for $\sigma^{a_n a_n}$ and $\sigma^{s_m s_m}$. 

\subsection{Numerical integration}

With the Eq.~(\ref{eq:HD1}) Hamiltonian, we integrate the evolution operator for the composite pulse,
\begin{equation}
U_{\rm c}({\bf r},\{\mathcal{E}_j,\varphi_j\})=\hat T e^{-i \int_0^{\tau_c} H_{\rm D1}({\bf r},t') dt'/\hbar}\label{eq:Uc}
\end{equation}
to propagate electronic state of stationary atom at location ${\bf r}$. We are particularly interested in high quality $|g\rangle-|a\rangle$ inversions~\cite{He2020a}, which is characterized by an average inversion efficiency \begin{equation}
f({\bf r})={\rm tr}(U_c({\bf r}) \rho^{(0)} U_c^{\dagger}({\bf r}){\bf 1}_a). \label{eq:ga}
\end{equation}
Here ${\bf 1}_a=\sum_{a,n} |a_n\rangle\langle a_n|$ is the projection operator into the 5P$_{1/2}$ ``$|a\rangle$'' manifold. We similarly define ${\bf 1}_g=\sum_m |g_m\rangle\langle g_m|$ and  ${\bf 1}_d=\sum_m |d_m\rangle\langle d_m|$. Here the initial atomic state is described by the density matrix $\rho^{(0)}=\frac{1}{7}{\bf 1}_g$ for $^{85}$Rb so that $\rho^{(0)}_{gg}=1$.

In addition, related to the experimental observation in this work is a normalized ground state depletion efficiency for an initially unpolarized atom, defined by Eq.~(\ref{eq:fg}) in the main text. The ground-state population after the $U_c$ control is evaluated as
\begin{equation}
    \rho_{gg}({\bf r},\tau_c)={\rm tr}(U_c({\bf r}) \rho^{(0)} U_c^{\dagger}({\bf r}){\bf 1}_g).\label{eq:rhougg}
\end{equation}
Here the initial atomic state is described by the density matrix $\rho^{(0)}=\frac{1}{12}({\bf 1}_g+{\bf 1}_d)$ for $^{85}$Rb so that $\rho_{gg}^{(0)}=7/12$.

Numerical evaluation of $f({\bf r}_{\perp})$, $f_g({\bf r}_{\perp})$ according to Eqs.~(\ref{eq:hyb}-\ref{eq:rhougg}), as those in Fig.~\ref{fig:Setup}, Fig.~\ref{fig:OnePulse}, Fig.~\ref{fig:ThreePulse} in the main text, are implemented in a straightforward manner with Matlab~\cite{Qiu2022}. 


\begin{figure}
    \centering
    \includegraphics[width=0.48\textwidth]{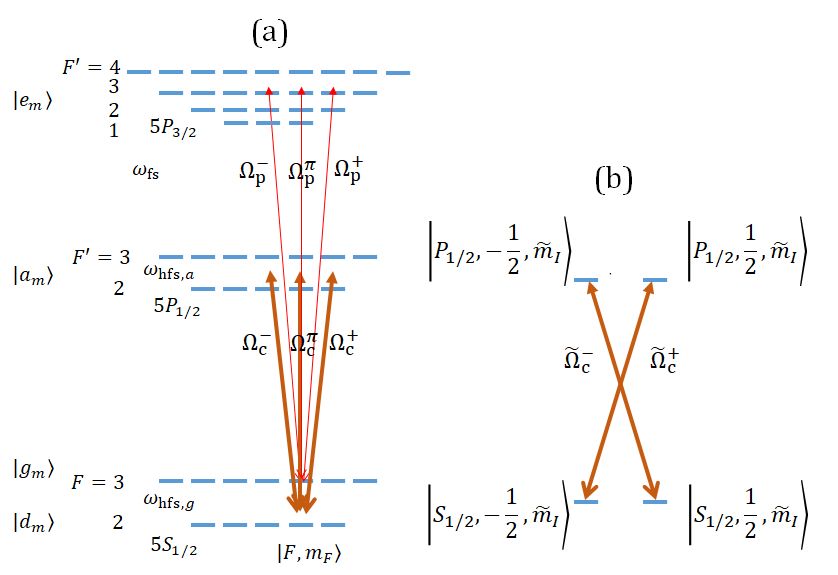}
    \caption{(a): The full level representation of the Fig.~\ref{fig:Setup}b scheme in the main text. With quantization axis along $z$, the local ${\bf E}_{\rm p,c}({\bf r})$ is decomposed into ${\bf e}_{\pm, z}$ directions to drive the $\Omega_{\rm p,c}^{\pm,\pi}$ hyperfine couplings respectively. The fine and hyperfine mixing lead to 2-photon Raman couplings among the ground-state hyperfine sublevels (not shown in the diagram). (b):  With $\tau_c \ll 1/\omega_{{\rm hfs},g},1/\omega_{{\rm hfs},a}$, the light-atom interaction is decomposed into 2-level $\sigma^{\pm}$ couplings in the $|nl_J, I, \tilde m_J,\tilde m_I\rangle$ basis ($n=5$, $I=5/2$ are omitted in the figure). The tilde signs emphasize the choice of quantization axis along the local helicity axis of light instead. The relative strength is determined by the ellipticity $\varepsilon({\bf r})$ (Eq.~(\ref{eq:omJI})). } 
\label{fig:coupling}
\end{figure}

\subsection{Composite control of hyperfine atom}\label{sec:JI}


Although the picosecond D1 interaction can be  evaluated numerically, the underlying physics can be obscured by the apparently complicated multi-level, multi-photon couplings. To understand the control robustness enabled by composite pulse techniques~\cite{Genov2014,Low2016}, a simpler picture of D1 transition dynamics without the hyperfine structure has been outlined in Appendix~\ref{sec:composite}. The picture remains valid for atoms with hyperfine splittings if the optical excitation is short enough: $\tau_c\ll 1/\omega_{{\rm hfs},g},1/\omega_{{\rm hfs},a}$. In this limit,  the light-atom interaction can be written in the $|nl_J,I, \tilde m_J,\tilde m_I\rangle$ basis with separately ``conserved'' electron and nuclear angular momenta ${\bf J}$ and ${\bf I}$ (Fig.~\ref{fig:coupling}b), with a local quantization axis along the helicity axis of the elliptical field (Eq.~(\ref{eq:eh})) for the HE$_{11}$ field ${\bf E}_c({\bf r})$. Since the hyperfine mixing of the $\tilde m_J+\tilde m_I=\tilde m_F$ levels are negligible during $\tau_c$, the Fig.~\ref{fig:coupling}b 2-level $\sigma^{\pm}$ transitions are decoupled like  Fig.~\ref{fig:composite}b. As such, the control of ``optical spin'' defined on any pair of $|g_m\rangle$ (or $|g_m\rangle$ superposition) and the corresponding $|a_m\rangle$ levels is decomposed into simultaneous SU(2) control of the $|nl_J,I, \pm 1/2, \tilde m_I\rangle \leftrightarrow |nl_J,I, \mp 1/2, \tilde m_I\rangle$ spins,   albeit with unequal $\tilde \Omega_c^{\pm}$ in general. Nevertheless, for small $\varepsilon$ so the $|\tilde \Omega^+/\tilde \Omega^-|$ ratio is moderate (Eq.~(\ref{eq:omJI})), then the  relative control errors can be suppressed, quite naturally, by the intensity-error-resilient composite pulse techniques~\cite{Genov2014,Low2016} as in Fig.~\ref{fig:composite}.

On the other hand, if the field polarization is close to be linear with $\varepsilon\ll1$, 
then the short pulse requirement for achieving efficient 2-level composite control can be relaxed to $\tau_c\ll 1/\omega_{{\rm hfs},a}$ alone, since the tensorial Raman couplings are suppressed in alkaline atoms~\cite{Happer1972}. This situation is particularly relevant when longer control pulses with $\tau_c\sim 1/\omega_{{\rm hfs},g}\ll 1/\omega_{{\rm hfs},a}$ are applied, such as for the $N=3$ composite pulses in Sec.~\ref{sec:three}.



Beyond the short pulse limits, the composite pulses may be tailored to manage the phase shifts associated with hyperfine splitting~\cite{He2020a} and to suppress multi-level dynamics~\cite{Qiu2022}, so as to effectively achieve few-level precise control on the hyperfine manifold. 



\section{Modeling the D2 absorption}\label{sec:D2}

Our goal in this section is to set up an ``exact'' model for predicting the attenuation of the nanosecond probe pulse in the ONF - atomic vapor setup outlined by Fig.~\ref{fig:Setup} and Fig.~\ref{fig:probe}. 


\subsection{The D2 Hamiltonian}

Similar to Eq.~(\ref{eq:Ec}), we consider the probe pulse with a near-field profile 
\begin{equation}
{\bf E}_{\rm p}({\bf r},t)={\bf E}({\bf r}) S_{\rm p}(t).\label{eq:Ep}
\end{equation} 
The spatial profile ${\bf E}({\bf r})$ is again evaluated according to Eq.~(\ref{eq:hyb}), but at an orthogonal incident polarization according to the Fig.~\ref{fig:probe}a setup ($\Theta\rightarrow \Theta+\pi$). The nanosecond temporal profile, as in Fig.~\ref{fig:probe}, is described by $S_{\rm p}(t)$ with $|S_{\rm p}(t)|\leq 1$.  The Rabi frequencies to drive the $|g\rangle-|e\rangle$ and $|d\rangle-|e\rangle$ transitions are written as
\begin{equation}
\Omega_{e_n s_m}^{l}({\bf r},t)=\langle e_{n}|{\bf E}_{\rm p}({\bf r},t) \cdot {\bf d}_{l}|s_m\rangle/\hbar
\end{equation}
for all the $s=g,d$ states and with $l=-1,0,1$ for the $\sigma^+$, $\pi$, and $\sigma^-$ transitions respectively.


The Hamiltonian during the D2 interaction under the rotating wave approximation is written as
\begin{equation}
\begin{aligned}
     H_{\rm D2}({\bf r},t)  = & \hbar\sum_{e}(\omega_{e}-\omega_{e 0})\sigma^{e_n e_n} + \\   &\hbar\sum_{s=g,d} (\omega_{s}-\omega_{g0})\sigma^{s_m s_m}+
    \\
&\frac{\hbar}{2}\sum_{s={g,d}}\sum_l\Omega^{l}_{e_n s_m}({\bf r}, t)\sigma^{e_n s_m} + {\rm h.c.}
        \end{aligned}\label{eq:HD2}
    \end{equation}
The notation follows the same conventions as those for Eq.~(\ref{eq:HD1}). The probe Rabi frequencies are accordingly written in the $\omega_{e0,g0}$ frame with the resonant frequency $\omega_{\rm p}=\omega_{e0,g0}$ canceled. With $\omega_{gd}=\omega_{{\rm hfs},g}\gg 1/\tau_{\rm p}$ for the nanosecond pulse duration $\tau_{\rm p}$, the $|d\rangle$ states are invisible to the probe pulse. We therefore effectively set $\Omega_{e_n d_m}^l=0$ to reduce the computational cost.

\subsection{Spontaneous emission}    

We account for spontaneous emission for both the D1 and D2 lines with a stochastic wavefunction method~\cite{   Dalibard1992,Carmichael1993, Dum1992}. For the purpose, six quantum jump operators associated with D1 and D2 emissions are introduced as 

\begin{equation}
    \begin{aligned}
    C_a^l&=\sqrt{\Gamma_{\rm D1}}\sum_{a,s=g,d} \mathcal{C}_{s_m,a_n}^l \sigma^{s_m a_n},\\
    C_e^l&=\sqrt{\Gamma_{\rm D2}}\sum_{e,s=g,d} \mathcal{C}_{s_m,e_n}^l \sigma^{s_m e_n}.
    \end{aligned}
\end{equation}
The $\mathcal{C}^l_{s,a}$ $\mathcal{C}^l_{s,e}$ are decided by the Clebsch-Gordan coefficients for the hyperfine transitions. As suggested in Sec.~\ref{sec:model} in the main text, the impact of surface interactions to the probe absorption are negligible within the nanosecond evolution. We therefore set $\Gamma_{\rm D1}=(27.7~{\rm ns})^{-1}$, $\Gamma_{\rm D2}=(26.2~{\rm ns})^{-1}$ as the natural linewidths of the free atom~\cite{Steck2019}.

The atomic density matrix $\rho(t)$ evolves according to the optical Bloch equation as
\begin{equation}
    i\hbar\dot\rho=[H_{\rm D2},\rho]-\frac{i\hbar}{2}\{\hat \Gamma,\rho\}+\hbar\sum_{l}C_a^l\rho C_a^{l \dagger}+C_e^l\rho C_e^{l\dagger}\label{eq:OBE}
\end{equation}
with
\begin{equation}
    \begin{aligned}
    &\hat\Gamma =\hat\Gamma_{\rm D1}+\hat\Gamma_{\rm D2}, {\rm with}\\
    &\hat \Gamma_{\rm D1}=\Gamma_{\rm D1}\sigma^{a_m a_m}\\
    &\hat \Gamma_{\rm D2}=\Gamma_{\rm D2}\sigma^{e_n e_n}
    \end{aligned}
\end{equation}
As to be detailed next, we evaluate Eq.~(\ref{eq:OBE}) by averaging the stochastic wavefunctions~\cite{Dum1992} with the near-field optical parameters decided by classical trajectories.

\subsection{Sampling the thermal atomic distribution}
As outlined in Sec.~\ref{sec:model} in the main text, we treat the center-of-mass motion of thermal atoms classically with prescribed ballistic trajectories ${\bf r}(t)={\bf r}_0 + {\bf v}_0 t$, which enter Eq.~(\ref{eq:OBE}) as time-dependent parameters. 

In the Monte Carlo simulation to be introduced next, the initial $({\bf r}_0,{\bf v}_0)$ is randomly sampled according to the phase-space distribution $g({\bf r}_0,{\bf v}_0)$ for the thermal atomic vapor uniformly surrounding the ONF as 
\begin{equation}
g({\bf r}_0,{\bf v}_0)= \sqrt{\left(\frac{M}{2\pi k_{\rm B}{\rm T}}\right)^3} \mu({\bf r}_0) e^{-M |{\bf v}_0|^2 /2 k_{\rm B}{\rm T}}.\label{eq:g}
\end{equation}
We set ${\rm T}=360$~K according to the in-vacuum thermometer readout. $M$ is atomic mass of $^{85}$Rb and  $k_{\rm B}$ is the Boltzmann constant. 
The stationary $g({\bf r},{\bf v})$ distribution is maintained by a detailed balance of microscopic transportation across the ONF near field. We ignore the impact of ONF surface and optical forces to $g({\bf r}_0,{\bf v}_0)$.

\subsection{Monte Carlo simulation}\label{sec:mc}

As in Fig.~\ref{fig:probe}d in the main text, we consider a cylindrical volume around ONF with a radius of $R=1~\mu$m and a length of $L=3$~mm to fully cover the near-field interaction.  Within the volume, we assume a uniform atomic density $\mu=P/k_{\rm B} {\rm T}$ maintained by mesoscopic thermal transportation with $P\approx 10^{-3}$~pascal. Numerically, individual atoms enter the volume through the $r_{\perp}=R$ surface
at random time. The atomic velocity, incident angle, and flux density obey the Maxwell's distribution at ${\rm T}=360$~K.  We choose the simulation time interval of $-t_{\rm w}<t<\tau_c+\Delta t+ t_{\rm p}$, with the control pulse at $0<t<\tau_c$ and probe pulse at $\tau_c+\Delta t< t<\tau_c+\Delta t+\tau_{\rm p}$. A $t_{\rm w}=30$~ns time window is chosen for the classical Monte Carlo trajectories to reach thermal equilibrium within the cylinder, before the control pulse is fired. 

To efficiently simulate the mesoscopic optical response including both the classical and quantum randomnesses, we sample many classical atomic trajectories ${\bf r}(t)={\bf r}_0+{\bf v}_0 t$ and evaluate a single stochastic wavefunction~\cite{Dum1992} for each ${\bf r}(t)$. The optical response is then evaluated by the ensemble average of expectation values. The numerical method is detailed as following.

The atomic initial state $|\psi(t=0)\rangle$ is set as one of the $\{|g_m\rangle,|d_m\rangle\}$ internal ground states. If the trajectory hits the nanofiber surface later, we assume the atom is immediately scattered back to the volume with a new random velocity according to the Maxwell velocity distribution, and with the internal state reset to one of the $\{|g_m\rangle,|d_m\rangle\}$ states. 

During $0<t<\tau_c$, we ignore atomic motion and evolve $|\psi(t)\rangle$ unitarily with $H_{\rm D1}({\bf r}_0,t)$, leading to
\begin{equation}
    |\psi(\tau_c)\rangle=U_c({\bf r}(0),\{\mathcal{E}_j,\varphi_j\})|\psi(0)\rangle
\end{equation} 
with $U_c$ from Eq.~(\ref{eq:Uc}). 

We now consider the evolution of the stochastic wavefunction, $|\psi_S(t)\rangle$, during $\tau_c<t<\tau_c+\Delta t+\tau_{\rm p}$ subjected to the $H_{\rm D2}$ Hamiltonian and $C_{a,e}^l$ quantum jumps. Here, taking advantage of the fact that the D1 excited states $|a\rangle$ are not affected by the D2 probe couplings, we reduce the simulation complexity by restricting $|\psi_S(t)\rangle$ within the D2 manifold, initiated at a ``quantum jump'' time $t^j_S$ which is the solution to~\cite{Dum1992}
\begin{equation}
    \rho_{aa}(\tau_c) e^{-\Gamma_a t^j_S}=r.\label{eq:tj}
\end{equation}
Here $r\in [0, 1]$ is a random number.  $\rho_{aa}(\tau_c)=\langle \psi(\tau_c)|{\bf 1}_a |\psi(\tau_c)\rangle$ is the atomic population in states $|a\rangle$ immediately after the control pulses. Given $\rho_{aa}(\tau_c)>r$ to guarantee a $t^j_S$ solution, the internal state for the D2 simulation is initialized as
\begin{equation}
|\psi_S(t^j_S)\rangle=C_a^l |\psi(\tau_c)\rangle,
\end{equation}
heralded by a $l-$polarized emission of D1 photon. The branching ratio for the $l=\sigma^-,\pi,\sigma^+$ emission is decided by the relative probabilities $p_l=\langle \psi(\tau_c)|C_a^{l\dagger}C_a^l|\psi(\tau_c)\rangle$. 

On the other hand, for $\rho_{aa}(\tau_c)<r$ so that there is no solution to $t^j_S$, then we set $t^j_S=\tau_c$ with atomic state simply being projected as
\begin{equation}
    |\psi_S(t^j_S)\rangle=({\bf 1}_g+{\bf 1}_a) |\psi(\tau_c)\rangle.
    \end{equation}

During $t^j_S<t<\tau_c+\Delta t+\tau_p$, the stochastic wavefunction $|\psi_S(t)\rangle$ evolves according to the effective Hamiltonian $H_{\rm eff}=H_{\rm D2}-i\hat \Gamma_{\rm D2}/2$ and is probabilistically interrupted by the D2 emission associated with $C_{e}^l$~\cite{Dum1992}.

Following Eq.~(\ref{eq:gamma1}) in the main text, the D2 probe scattering rate is numerically evaluated as 
\begin{equation}
        \bar \gamma(t)=\frac{1}{N_S}\sum_{S=1}^{N_s} \theta(t,t^j_S) {\rm Im}\left[\sum_{e,l}   \langle \psi_S|e_m \rangle  \langle g_n| \psi_S\rangle \Omega_{e_m g_n}^l\right] \label{eq:gm2}
\end{equation}
with normalized state vectors $| \psi_S(t)\rangle$. The step function $\theta(t,t_S^j)$ is 1 for $t>t_S^j$, and 0 otherwise. By removing the control evolution, straightforward simplification of Eq.~(\ref{eq:gm2}) can evaluate  the scattering rate $\bar\gamma_0(t)$ for the atomic vapor in absence of the control pulses. For a particular experimental configuration, $\bar\gamma(t)$, $\bar\gamma_0(t)$ are typically  evaluated with $N_{S}=10^6$ trajectories, each takes about 1~h time on a PC cluster (Intel-i7 34 cores). The probe absorption $A$ and the normalized difference $\overline{\delta T}$ are then evaluated according to Eq.~(\ref{eq:deltaTt}) in the main text.  

\begin{figure}
    \centering
    \includegraphics[width=0.45\textwidth]{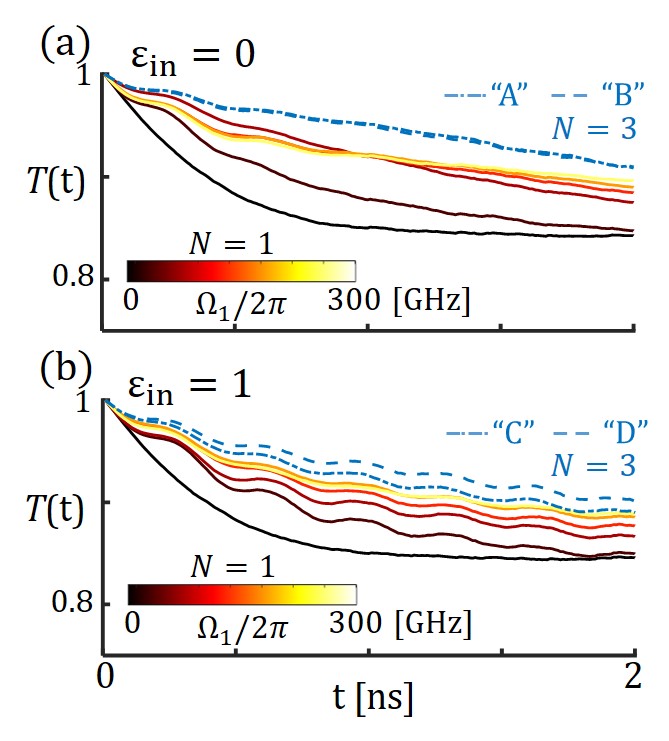}
    \caption{Monte Carlo simulation of the transient transmission $T(t)$ during the $\tau_{\rm p}=2$~ns probe (Also see Fig.~\ref{fig:probe}(b)(c)). Figs.~(a)(b) are with linear ($\varepsilon_{\rm in}=0$) and circular ($\varepsilon_{\rm in}=1$) HE$_{11}$ incidences respectively. For $N=1$ control, the pulse energy is $\mathcal{E}_1= 0,0.01,0.05,0.15,0.3,0.6$~pJ, associated with peak Rabi frequencies $\Omega_1= 2\pi\times 0,30,75,125,180,255$~GHz (associated with Fig.~\ref{fig:OnePulse}b in the main text). For $N=3$ control, the phase combination is set according to the marks ``A''-``D'' in Fig.~\ref{fig:ThreePulse}(h)(i) in the main text. The pulse energy are $\mathcal{E}_j=0.05$~pJ (``A'',``B'') and $\mathcal{E}_j=0.02$~pJ (``C'',``D'') respectively with the associated $\Omega_j= 2\pi\times 75,50$~GHz.} 
\label{fig:probetransient}
\end{figure}

The Monte Carlo simulation enables us to look into nanosecond absorption dynamics during the nanosecond probe itself (Fig.~\ref{fig:probe}b,c) ,  which is not resolved experimentally (Appendix~\ref{sec:electronics}). As in Fig.~\ref{fig:probetransient} numerical example for the linear $\varepsilon_{\rm in}=0$ and circular $\varepsilon_{\rm in}=1$ HE$_{11}$ couplings, in absence of the picosecond control pulses (black lines), the probe transmission $T(t)$ always relax from $T(\tau_c)=1$ to the steady-state $T_{\rm ss}\approx 85\%$ in this example within one nanosecond. Physically, the initial $T=1$ is associated with the zero atomic dipole $\langle {\bf d}\rangle$ initially, which is then driven by ${\bf E}_{\rm p}$ in a time-dependent, off-resonant fashion due to the 100~MHz-level Doppler shift and nanosecond transient associated with the atomic motion along $z$ and $(x,y)$ respectively. Thermal ensemble average of the oscillatory scatterings, as by Eq.~(\ref{eq:gamma1}), leads to the rapid relaxation of $T$ to $T_{\rm ss}$ within a nanosecond. The relaxation time is predominantly decided by the $2\sqrt{2{\rm ln}2}\beta_{\rm p}\sqrt{k_B {\rm T}/M}\approx 2\pi\times 700$~MHz Doppler width and is further shortened by the transient broadening of the mesoscopic gas.  Indeed, the choice of $\tau_{\rm p}=2~$ns in this work was made to balance the noise level with the rapid decay of the transient absorption.

On the other hand, as the colored solid-curve examples in Fig.~\ref{fig:probetransient}, after subjecting an $N=1$ picosecond control ($\Delta t=0$), the probe transmission $T(t)$ decreases much slower.  Physically, the picosecond excitation breaks the atomic distribution from the Eq.~(\ref{eq:g}) thermal equilibrium, and it takes time for ``new'' atoms in $|g\rangle$ to transport to the interface, albeit with a random ``time of arrival'' microscopically and therefore with fast dipole transients being self-suppressed on average.  
Similar relaxations are found in the simulation for the $N=3$ composite excitations given by the dashed lines in Fig.~\ref{fig:probetransient}(a)(b), where the simulation parameters are according to the ``A''-``D'' combinations in the Fig.~\ref{fig:ThreePulse} measurements. 

Similar to the slowed relaxation in the Fig.~\ref{fig:probetransient} curves of $T(t)$ subjected to the control pulses, for the $\tau_{\rm p}$-integrated probe absorption (the normalized transient transmission $\overline{\delta T}$ by Eq.~(\ref{eq:deltaT})) in the $\Delta t$-delayed control-probe measurements (Fig.~\ref{fig:probe}e), the recovery to the steady-state value is also decided by the refilling of ground state atoms into the ONF near-field region, primarily by the mesoscopic transportation from far away. With in mind the exponential form of the evanescent tail (Fig.~\ref{fig:Setup}), it is easy to show that the ``refilling time constant'' depends logarithmically on the control strength $\Omega_{\rm c}$. We note the hyperfine Rabi frequency is itself polarization dependent (Fig.~\ref{fig:composite}b, Fig.~\ref{fig:coupling}). As a result, the recovery dynamics is slightly more complex in the $\varepsilon_{\rm in}=0$ linear incidence case (Fig.~\ref{fig:probetransient}a) due to the more strongly varying near-field $\varepsilon(x,y)$ (Fig.~\ref{fig:Setup}c). 
By setting $\mathcal{E}_1=0.5$~pJ in the simulation according to the Fig.~\ref{fig:probe}e measurements ($\varepsilon_{\rm in}=0$) and by rescaling the experimental $\overline{\delta T}$ as explained in Sec.~\ref{sec:calib}, fairly good agreement is obtained between the experimental and numerical $\Delta t$-$\overline{\delta T}$ data  Fig.~\ref{fig:probe}e.

Notably, in Fig.~\ref{fig:probetransient}(a)(b) we see weak modulation in all the $T(t)$ curves with $\sim 0.3$~ns periodicity, a subtle interference between the transient ${\bf d}_{e g}$ and ${\bf d}_{e d}$ dipoles. 
As discussed in Sec.~\ref{sec:three}, the ${\bf E}_{\rm c}$ picosecond control is long enough to induce $|g\rangle - |d\rangle$ hyperfine excitation, particularly if the optical polarization is circular. On the other hand, the rapidly rising nanosecond probe pulse, even though with $\omega_{\rm p}$ centered to the $\omega_{eg}$ hyperfine transition (Fig.~\ref{fig:Setup}b, Fig.~\ref{fig:coupling}a), has enough spectrum component to off-resonantly excite the $|d\rangle-|e\rangle$ transition. The ``quantum beat'' in all the Fig.~\ref{fig:probetransient} curves is a result of the $|d\rangle-|g\rangle$ Raman coherence induced by the picosecond control pulse, so that the tiny ${\bf d}_{ed}$ dipole is phased to interfere with the ${\bf d}_{eg}$ dipole at $\omega_{{\rm hfs},g}=2\pi\times 3.04$~GHz. The beat note is more pronounced for circular HE$_{11}$ excitation with $\varepsilon_{\rm in}=1$, as expected according to Sec.~\ref{sec:three}. Nevertheless, these $\omega_{{\rm hfs},g}$-scale optical transients average out in the $\delta T$ measurements with $\tau_{\rm p}$-integration. The rapid average of coherent transients supports a much simpler diffusive average model, to be discussed in the following.

\subsection{The diffusive average model}\label{sec:conv}

The diffusive average model by Eq.~(\ref{eq:dT2}) in the main text starts with calculating $f_g({\bf r})$ according to Eqs.~(\ref{eq:rhougg})(\ref{eq:fg}) and $i_{\rm p}=|{\bf E}_{\rm p}|^2$ according to Eqs.~(\ref{eq:Ep})(\ref{eq:hyb}). The model is based on the observation that for atom with velocity ${\bf v}$ and location ${\bf r}(t)$, its contribution to the overall photon scattering rate (Eq.~(\ref{eq:gamma1}) in the main text) is on-average proportional to the population $\rho_{gg}(t)$ and the probe intensity $i_{\rm p} ({\bf r}(t))$ in the linear probe regime, where we further have $\rho_{gg}(t>\tau_{\rm c})\approx \rho_{gg}(\tau_c)$ within the nanosecond $\tau_{\rm p}$ time. 
For thermal atoms uniformly sampling the phase space, we expect errors associated with the coherence transients to average out quite efficiently. The residual transients that survive the average, as those exemplified in Fig.~\ref{fig:probetransient}, have additional chance to be suppressed by the $\tau_{\rm p}$-average and then cancel each other in the $\overline{\delta T}$ ratio (Eq.~(\ref{eq:deltaTt})). Finally, since the ``impulse'' excitation hardly change the atomic velocity distribution, the reductions of average atomic absorption by the Doppler and transient broadening are largely shared by the unperturbed and the excited vapors. The effects are thus expected to be largely cancelled in the $\overline{\delta T}$ ratio (Eq.~(\ref{eq:deltaTt})) too. 

\begin{figure}
    \centering
    \includegraphics[width=0.45\textwidth]{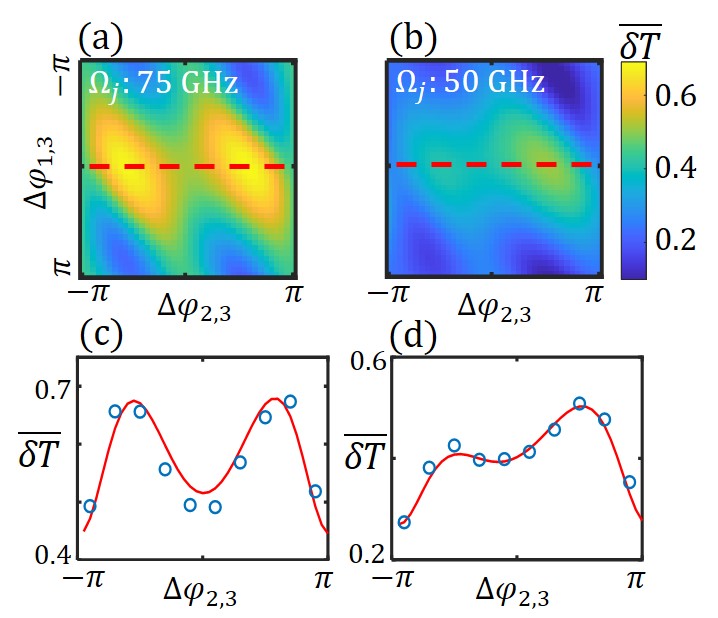}
    \caption{Comparison of the $\overline{\delta T}$ simulations based on the diffusive average (Eq.~(\ref{eq:deltaTt})) and the Monte Carlo method (Eq.~(\ref{eq:dT2})). Here (a,b) are the same simulation data as those in Fig.~\ref{fig:ThreePulse}(h,i). The solid lines in (c,d) are extracted from (a,b) along the $\Delta\varphi_{1,3}=0$ dashed lines. The circles in (c,d) are from the Monte Carlo simulations.} 
\label{fig:compare}
\end{figure}

To confirm the validity of the Eq.~(\ref{eq:dT2}) approximation we repeat the Fig.~\ref{fig:ThreePulse}(h)(i) calculation along the $\Delta \varphi_{1,3}=0$ line using the Monte Carlo method (Eq.~(\ref{eq:deltaTt})). The results are presented in Fig.~\ref{fig:compare}(c)(d) with circular symbols. By adjusting the diffusive length to be $\xi_{\rm p}'=0.9(\sqrt{2}v_{\rm T}\tau_{\rm p}/2)$ (with $\sqrt{2}$ to account for the 2D motion), the difference of $\overline{\delta T}$ between the predictions by the ``exact'' and diffusive average models is typically less than 0.05.

\section{Experimental Detail}

\subsection{Pulse sequence generation system}\label{sec:generator}

\begin{figure*}
        \centering
        \includegraphics[width=0.8\textwidth]{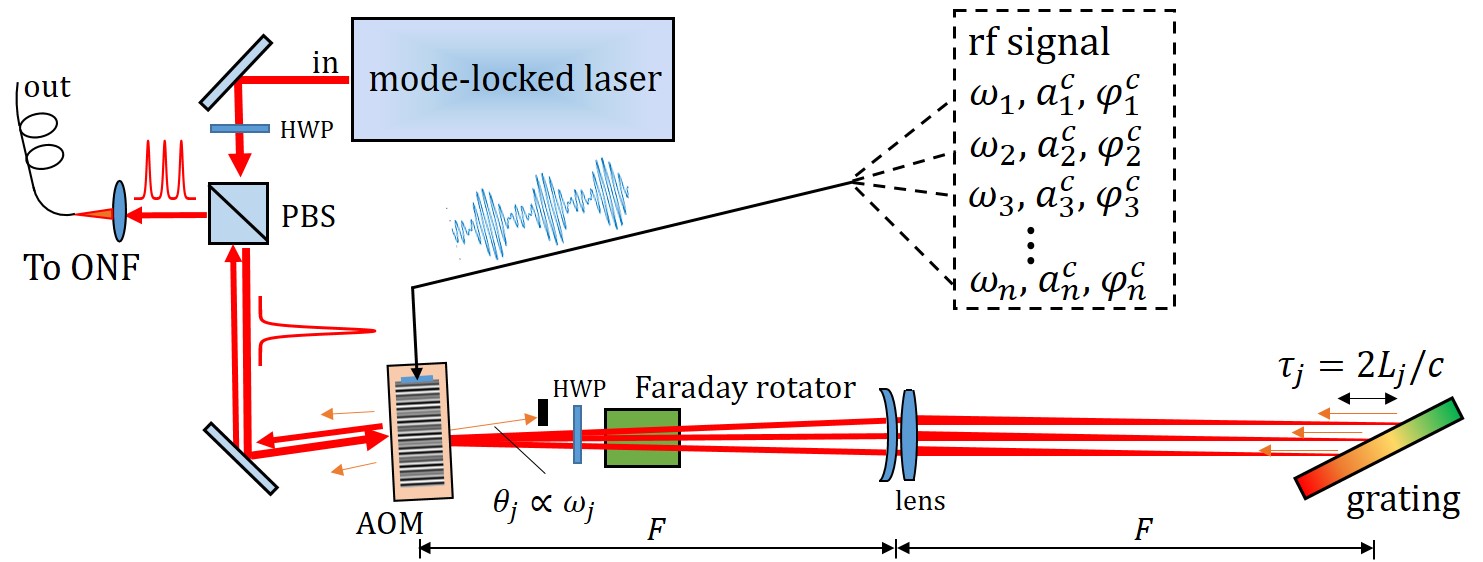}
        \caption{Schematic of the picosecond pulse sequence generation system~\cite{Ma2020} for the ONF-interface atomic state control in this work. A transform-limited picosecond pulse is diffracted by a multi-frequency-driven acousto-optical modulator (AOM) into multiple paths, retro-diffracted by a grating to double-pass the AOM with tunable delay $\{\tau_j\}$. The output pulses along the time-reversed direction is separated from the input using a polarization beam splitter (PBS), repetition rate prescaled and pulsed picked (not shown), before being coupled to a single-mode fiber toward the nanofiber experiment. Here $\tau_j=(j-1) \tau_d$  with $\tau_d=24$~ps. The amplitude and phase of each pulse, $\{a_j,\varphi_j\}$, is programmed by the amplitude and phase of radio-frequency (rf) sideband signals driving the AOM.} 
    \label{fig:generation}
\end{figure*}

Our nanofiber-atom interface technique relies on coherent generation of composite sequence of picosecond pulses with precisely tunable amplitude and phase, $\{a_j,\varphi_j\}$, to optimize the atomic state contrl. In this work, the composite picosecond pulse generator is based on a time-domain pulse shaping method developed recently~\cite{Ma2020}, as schematically illustrated in Fig.~\ref{fig:generation} and briefly summarized in the following. 

We use a Ti-Sapphire mode-locked laser (Spectra-Physics Tsunami system) to generate  transform-limited picosecond pulses with $\tau_0=12~$ps at a repetition rate of $f^{(0)}_{\rm rep}=80$~MHz. The pulsed output, referred to as ${\bf E}_{\rm in}(t)$ in the following, is directed to a multi-frequency $\{\omega_j\}$-driven double-pass acousto-optical modulator (AOM). Instead of retro-reflecting the multi-diffracted beams, we use a high-density grating (2400 line/mm) to retro-diffract each path backward for the second AOM diffraction. The grating retro-diffraction introduces a path-dependent delay $\tau_j=2 L_j/c\propto\omega_j-\omega_1$ relative to $\tau_1$. Among the AOM double-diffracted beams, the direction-reversed beam is picked by a polarization beamsplitter, repetition rate pre-scaled~\cite{liu2022} and pulse-picked (not shown in Fig.~\ref{fig:generation}) to $f_{\rm rep}=2$~MHz, single-mode selected, before being combined with the probe and coupled into ONF (Fig.~\ref{fig:Setup}).  In this work, the tunable delay for $\tau_j=(j-1)\tau_d$ is set by $\tau_d=24$~ps. Taking into account an overall loss coefficient $\kappa$, the shaped composite pulse output, as a sum of individually delayed sub-pulse ${\bf E}_{j,\rm out}(t)$, has a complex envelop function~\cite{Ma2020}
\begin{equation}
    \begin{aligned}
{\bf E}_{\rm c}(t)&=\kappa \sum_j^N 
    {\bf E}_{j,\rm out}(t),\\
    &=\sum_j^N a_j e^{i \varphi_j} {\bf E}_{\rm in}(t-\tau_j)
\end{aligned}\label{equ:inOut}
\end{equation}
at the ONF interface. The amplitudes and phases of the composite pulses $\{a_j,\varphi_j\}$ are controlled by that of the $f_{\rm rep}$-synchronized rf waveforms $\{a^c_j,\varphi^c_j\}$ as $a_j\propto \kappa (a_j^{\rm c})^2$ and  $\varphi_j=2\varphi_j^{\rm c}+$constant for the weakly driven AOM, as described in ref.~\cite{Ma2020}. With the multiple beams sharing a common optical path, the composite pulse shaped by the rf-programmed multi-frequency AOM is amplitude and phase stable over (many) hours. Inspired by related work for ``direct space-to-time pulse shaping''~\cite{Emplit92, Leaird99, Mansuryan2011}, we refer this method as ``direct ${\bf k}$-space-to-time pulse shaping''.

\subsection{High speed signal acquisition and averaging}\label{sec:electronics}
\begin{figure}
    \centering
    \includegraphics[width=0.48\textwidth]{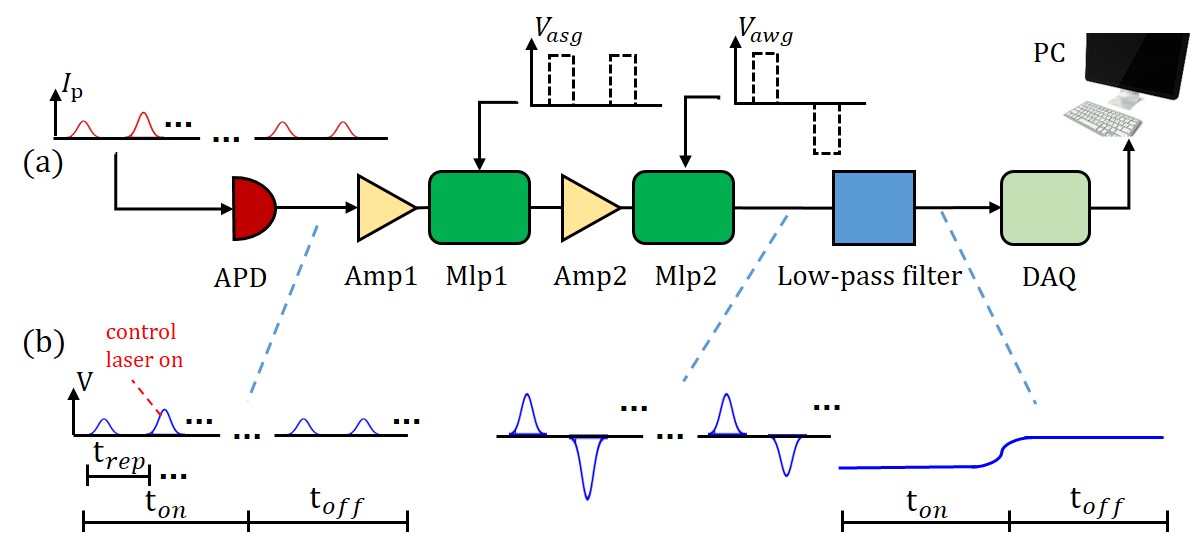}
    \caption{High speed signal acquisition and averaging. (a): The schematic setup, with the processed signals at each stages represented by the  blue curves in (b), from the nanosecond voltage pulse pairs at the APD output, to the slowly varying voltage levels sampled by the DAQ card. The synchronized signals to control the two multipliers (Mlp1 and Mlp2) are generated by an arbitrary sequence generator (CIQTEK ASG8000) and an arbitrary waveform generator (CIQTEK AWG4100) respectively. The control pulse is combined with every other probe pulse for $t_{\rm on}=1.25$~ms and then turned off for $t_{\rm off}=1.25$~ms for removing the voltage offset. }
    \label{fig:Acquisition}
\end{figure}

In the control-probe experiment, we keep the peak power of the nanosecond D2 probe at $P_{\rm p}\approx 10$~nW level to avoid saturation. The signal is close to the noise-equivalent power of the 1 GHz avalanche photodiode (APD) module (Hamamatsu C5658). Fortunately, taking advantage of rapid probe repetition at 4~MHz, 
enabled by the rapid recovery of the thermal vapor-ONF system (Fig.~\ref{fig:probe}e), it is still possible to retrieve transient transmission $\delta T$ with a $0.1\%$ sensitivity within seconds by rapidly averaging the difference of transmission induced by the control pulses. 

We use off-the-shelf rf components to construct a signal averager schematically illustrated in Fig.~\ref{fig:Acquisition}. The $\tau_{\rm p}=2~$ns probe pulse is sent to probe the ONF system with a  $t_{\rm rep}'=250$~ns repetition time. A synchronized control pulse is fired immediately before every other probe pulse to form a $\delta T$ measurement cycle with the nanosecond pulse pairs. As outlined in Fig.~\ref{fig:Acquisition}, after the APD module receive the pulse pairs, they are amplified (Mini-Circuits ZFL-500+) and subjected to two multipliers (AD834) for time-domain windowing and pulse sign reversals. The processed signals are then averaged by a 10 kHz bandwidth low-pass filter. The integrated signal level $\delta V$ reflects the difference of nanosecond probe transmission induced by the picosecond control. The control pulse is combined with every other probe pulse for $t_{\rm on}=1.25$~ms and then turned off for $t_{\rm off}=1.25$~ms for removing the voltage offset. The alternating measurement and calibration cycles ensure any slowly varying electronic offset is removed. To compare $\delta V$ with the probe signal level itself, we remove the second pulse of the probe pulse pair during a $t_0=20$~ms interval of $t_{\rm int}=200$~ms integration cycle to record $V_0$.  The low-passed signal is send to computer through a data acquisition (DAQ) card (NI USB-6363). We integrate $10^7$ differential measurements in two seconds to obtain a $\delta T=\delta V/V_0$ readout. The measurement is remarkably accurate with a $0.1\%$ sensitivity, which is inferred from the rms deviation from repeated $\delta T$ readouts.

\subsection{HE$_{11}$ mode polarization control and measurement}\label{sec:pol}
\begin{figure}
    \centering
    \includegraphics[width=0.45\textwidth]{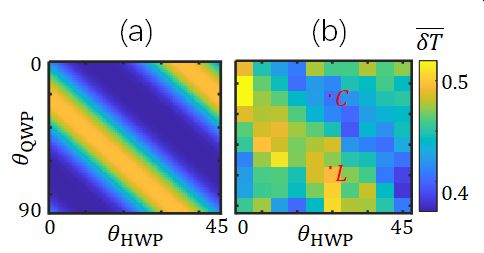}
    \caption{Simulated (a) (Eq.~(\ref{eq:dT2})) and experimentally measured (b) $\overline{\delta T}$ at $\mathcal{E}_1=0.2$~pJ (Appendix~\ref{sec:calib}), when the polarizations of the orthogonally polarized control and probe pulses (Fig.~\ref{fig:probe}a) are scanned by the HWP and QWP plates before they are coupled to the HE$_{11}$ mode of ONF. Two angular combinations for realizing linear and circular polarized HE$_{11}$ modes are marked with ``C'' and ``L'' in the experimental data graph in (b). }
    \label{fig:PolarScan}
\end{figure}


We use a pair of automated half-wave plate (HWP) and quarter-wave plate (QWP) in front of the nanofiber coupler (Fig.~\ref{fig:probe}a) to control the polarization states of the HE$_{11}$ mode, for the orthogonally polarized control and probe pulses. An example polarization-dependent $\overline{\delta T}$ measurement with a 2D polarization scan is shown in Fig.~\ref{fig:PolarScan}. As suggested by Fig.~\ref{fig:OnePulse} of the main text and according to the numerical simulations, the polarization dependence of the transient probe transmission with single control pulses is most pronounced near $\mathcal{E}_1\sim $0.2~pJ, which is also the pulse energy value set in this measurement. Representative waveplate angular combinations with $\varepsilon_{\rm in}=0$ and $\varepsilon_{\rm in}=1$ are marked with  ``L'' and ``C'' respectively. The angular combinations are sampled in picosecond controls with single and composite pulses as those in Fig.~\ref{fig:OnePulse} and Fig.~\ref{fig:ThreePulse}. 




\subsection{$A$ and $\mathcal{E}_j$ estimations in presence of slow drifts }\label{sec:calib}


To obtain normalized $\overline{\delta T}$ as those in Figs.~\ref{fig:probe}~\ref{fig:OnePulse}~\ref{fig:ThreePulse}~\ref{fig:PolarScan} from the transient transmission $\delta T$ measurements as outlined in Appendix~\ref{sec:electronics}, the atomic absorption $A$ across ONF (Fig.~\ref{fig:probe}b) needs to be accurately measured for the normalization (Eq.~(\ref{eq:deltaT})). In addition, to compare the experimental $\overline{\delta T}$ with simulations, we need to estimate individual pulse energy $\mathcal{E}_j$ and the peak Rabi frequency $\Omega_j$ for the composite pulses. In this work, both $A$ and $\mathcal{E}_j$ are estimated with moderate accuracies. In the following we detail the procedure to measure these parameters and to rescale their peak values against unknown offsets.


As outlined in the main text, we record the ``steady state'' absorption $A$, {\it i.e.}, the nanosecond probe absorption by the mesoscopic vapor in equilibrium in absence of the control pulse excitation, with CW absorption spectroscopy~\cite{foot:slow}. For all the experiments in this work, the $A$ recordings were made only at the beginning of the control-probe measurements, due to technical reasons.  Fortunately, taking advantage of the simplicity in the single-pulse absorption depletion measurement as those in Fig.~\ref{fig:OnePulse}, we are able to infer $A$ in situ, using the approximate $A\approx 2\delta T$ relation from the $\delta T$ measurement itself at strong enough $\mathcal{E}_1$-excitations (Sec.~\ref{sec:sat}). By analyzing multiple data sets taken during different periods of this project, we conclude that during the initial hour of the rubidium dispenser operation, the ONF local vapor pressure could vary substantially to affect $A$. The slow pressure relaxation is likely associated  with surface atomic adsorption across the vacuum chamber. After the initial hour, the vapor pressure tends to stabilize for hours, during which most of the data presented in this work were taken. The stable measurement condition supports faithful retrieval of $\overline{\delta T}$ features as a function of the control parameters, as those in Figs.~\ref{fig:probe}~\ref{fig:OnePulse}~\ref{fig:ThreePulse}~\ref{fig:PolarScan}. However, we expect the absolute value of $A$ to be scaled by the unknown pressure drift after hours since the initial measurement. To counter the effect we hand-rescale $A$ as detailed below to best match the numerical simulations.  






To estimate the control pulse parameters $\Omega_j$, we measure the incident power $P$ of the control pulses with a calibrated power meter (Thorlabs PM160), before the pulses are coupled to ONF (Fig.~\ref{fig:generation}). Taking into account the pre-calibrated fiber coupling efficiency $\eta_C\approx 50\%$, the pulse energy $\mathcal{E}_j$ is estimated as $\mathcal{E}_j=\eta_C t_{\rm rep} P /N$ for the equal-amplitude $N-$pulse sequence. Here $t_{\rm rep}=500$~ns is the period of the pulse picking.  Automatic adjustments of the pulse energy is achieved by scanning the rf signal amplitudes $a_j^{\rm c}$ in the pulse generation system, which are pre-calibrated to the output pulse power. 


We define $\Omega_j$ as the peak Rabi frequency of the picosecond pulse at the nanofiber surface averaging over the angular direction $\phi$,
$\Omega_j\equiv a_j \langle (\Omega_{\rm c}({\bf r}))_{r=d/2}\rangle_{\phi}$. The $\Omega_j$ is estimated according to Eqs.~(\ref{eq:Ec})(\ref{eq:rRbi}), with ${\bf E}_c({\bf r},t)$ inferred from
\begin{equation}
    \frac{1}{2}\int \varepsilon_0   \mathcal{N}^2  |{\bf E_{\rm c}({\bf r},t)}|^2 v_c {\rm d}^2{\bf r}_{\perp} {\rm d}t=N \mathcal{E}_j. \label{eq:Energy1}
 \end{equation}
 Here $v_c=\omega_c/\beta_c$ is the phase velocity of the ONF guided control pulses ($\beta_c$ is the associated propagation constant) and $\mathcal{N}({\bf r}_{\perp})$ is the transverse ONF refractive index profile. 
 
 We denote the pulse energy $\mathcal{E}_j$ from the estimation above as $\mathcal{E}_{j,{\rm raw}}$. To simulate the experiments,  in the simulations we assume accurately programmed control pulse phase $\varphi_j$ and pulse interval $\tau_d$ according to Eq.~(\ref{eq:cPulse}). By rescaling $\mathcal{E}_{j,{\rm raw}}$ with a common factor $\eta_{\rm P}$, the experimentally observed $\overline{\delta T}$ features can be reproduced as those in Fig.~\ref{fig:probe}~\ref{fig:OnePulse}~\ref{fig:ThreePulse},~\ref{fig:PolarScan} with $\Omega_j=\sqrt{\eta_{\rm P}}\Omega_{j,{\rm raw}}$. For each graph in these figures, we typically find $\eta_{\rm P}\approx 1.3\sim1.6$ which is centered around $\eta_{\rm P}=1.45$. This level of systematic offset is expected, considering the moderate accuracy in the optical power estimation. On the other hand, the $\pm 10\%$ uncertainty around the mean $\eta_{P}$, applied to different graphs of Figs.~\ref{fig:probe}~\ref{fig:OnePulse}~\ref{fig:ThreePulse},~\ref{fig:PolarScan}, is illustrated in Fig.~\ref{fig:OnePulse}b with the horizontal double-sided arrow. 
 
 Finally, while the simulated $\overline{\delta T}$ patterns are matched to the measurements by the $\Omega_j$ adjustment, the simulation typically suggests  ``actual'' $\overline{\delta T}$ to be different slightly from the $\overline{\delta T}_{\rm raw}$ ``raw data''. The difference is highly likely due to the aforementioned rubidium pressure drifts. We therefore also assign a rescaling factor $\eta_{\delta T}$ to $\overline{\delta T}_{\rm raw}$, leading to $\overline{\delta T}=\eta_{\delta T} \overline{\delta T}_{\rm raw}$ in the figures  to best match the numerical $\overline{\delta T}$ values. The $\eta_{\delta T}$ for Fig.~\ref{fig:probe}, Fig.~\ref{fig:OnePulse}, Fig.~\ref{fig:ThreePulse}(j-m), Fig.~\ref{fig:ThreePulse}(n) and Fig.~\ref{fig:PolarScan} are $\{0.53,1.07,0.82,1.05,1.07\}$ respectively (The Fig.~\ref{fig:ThreePulse}(j-m) and Fig.~\ref{fig:ThreePulse}(n)  data were taken on different days.). The adjustments are therefore within $\pm 20\%$ as suggested by the vertical double-sided arrow in Fig.~\ref{fig:OnePulse}d, except for the Fig.~\ref{fig:probe}e data which is likely related to its early measurement time within the initial hour of the rubidium dispenser operation. We note that despite the uncertainty associated with the slow drifts, the consistent match between groups of measurements, such as  the Fig.~\ref{fig:OnePulse}b single pulse data, and the Fig.~\ref{fig:ThreePulse}(e-h) simulation and the uniformly down-scaled Fig.~\ref{fig:ThreePulse}(j-m) experimental data,  all strongly suggest close-to-ideal performance of the composite picosecond control in this work. In particular, an optimal absolute $\overline{\delta T}\approx 70\%$ is likely reached in the Fig.~\ref{fig:ThreePulse}(m) data, assuming $\overline{\delta T}\approx 50\%$ at $\Delta\varphi_{1,3}=\Delta\varphi_{2,3}=0$, as by Fig.~\ref{fig:compare}, similar to the single-pulse saturation effect discussed in Sec.~\ref{sec:sat}.

\bibliography{nfc}

\begin{thebibliography}{93}
\bibitem [{foo({\natexlab{a}})}]{foot:cs}%
  \BibitemOpen
  \bibinfo {note} {Associated with $T(t)-$relaxation as in
  Fig.~\ref{fig:probetransient}, the $\tau_p$-integrated $\overline{\delta T}$
  decreases slightly from $50\%$ with increased $\tau_{\rm p}$. However, with
  strong enough $\mathcal{E}_1$ to deplete atoms far away from ONF, the
  recovery is slowed enough that $\overline{\delta T}\approx 50\%$ for
  $\tau_{\rm p}=2$~ns is nevertheless reached with the pJ-level
  $\mathcal{E}_1$, as in Fig.~\ref{fig:OnePulse} according to
  Eqs.~(\ref{eq:deltaTt})(\ref{eq:dT2}) simulations.}\BibitemShut {Stop}%
\bibitem [{foo({\natexlab{b}})}]{foot:shaper}%
  \BibitemOpen
  \bibinfo {note} {An improved version of pulse shaper is developed with
  reduced parameter cross-talk to enable picosecond composite control with
  $N\geq 5$. The results will be given in a separate publication.}\BibitemShut
  {Stop}%
\bibitem [{foo({\natexlab{c}})}]{foot:slow}%
  \BibitemOpen
  \bibinfo {note} {To apply the slow-detector recorded probe absorption $A_0$
  to the nanosecond probe, a $\eta=0.82$ factor is multiplied as $A=\eta A_0$,
  which accounts for the reduced absorption for the nanosecond probe relative
  to the cw probe by the evanescently coupled atomic vapor, as unveiled by the
  Monte Carlo simulations (Fig.~~\ref{fig:probetransient}).}\BibitemShut
  {Stop}%
\end{thebibliography}

\end{document}